\documentstyle[preprint,aps,pre,eqsecnum]{revtex}
\begin{document}
\draft
\title{Phase Behavior of Type-II Superconductors with Quenched Point
Pinning Disorder: A Phenomenological Proposal}
\author{Gautam I. Menon\footnote{Email:menon@imsc.ernet.in}}
\address{The Institute of Mathematical Sciences, C.I.T. Campus, \\
Taramani, Chennai 600 113, India}
\date{\today}
\maketitle
\begin{abstract}

A general phenomenology for phase behaviour in the
mixed phase of type-II superconductors with weak point
pinning disorder is outlined.  We propose that the
``Bragg glass'' phase generically transforms via two
separate thermodynamic phase transitions into a
disordered liquid on increasing the temperature. The
first transition is into a glassy phase, topologically
disordered at the largest length scales; current
evidence suggests that it lacks the long-ranged phase
correlations expected of a ``vortex glass''.  This
phase has a significant degree of short-ranged
translational order, unlike the disordered liquid, but
no quasi-long range order, in contrast to the Bragg
glass.  This glassy phase, which we call a
``multi-domain glass'', is confined to a narrow sliver
at intermediate fields, but broadens out both for much
larger and much smaller field values. The multi-domain
glass may be a ``hexatic glass''; alternatively, its
glassy properties may originate in the replica
symmetry breaking envisaged in recent theories of the
structural glass transition.  Estimates for
translational correlation lengths in the multi-domain
glass indicate that they can be far larger than the
interline spacing for weak disorder, suggesting a
plausible mechanism by which signals of a two-step
transition can be obscured.  Calculations of the Bragg
glass-multi-domain glass and the multi-domain
glass-disordered liquid phase boundaries are presented
and compared to experimental data.  We argue that
these proposals provide a unified picture of the
available experimental data on both high-T$_c$ and
low-T$_c$ materials, simulations and current
theoretical understanding.

\end{abstract}
\vspace*{1cm}
\section{Introduction}\label{sec:intro}

In the mean-field phase diagram of a pure type-II superconductor,
Meissner, mixed and normal phases are separated by continuous phase
transitions associated with lower (H$_{c1}$(T)) and upper (H$_{c2}$(T))
critical fields \cite{review1}.  In the mixed phase, an applied
magnetic field $H$ enters the sample in the form of singly quantized
lines of magnetic flux.  These lines repel each other, stabilizing a
triangular line lattice.  Thermal fluctuations melt this lattice via a
first-order melting transition, subdividing the domain of the mixed
phase into solid and fluid phases of flux
lines\cite{nelson,houghton,us}.

With ${\vec H} = H{\hat z}$, flux-lines in the lattice 
phase are parametrized by two-dimensional 
coordinates ${\bf R}_i$, where ${\bf R}_i$ 
describes the position
of the i$^{th}$ line.  Deviations from this arrangement define a
two-component displacement field ${\bf u}({\bf R}_i,z)$, with
an associated coarse-grained elastic free energy cost
\begin{equation}
F_{el} = \int d{\bf r}_\perp dz~\large[\frac{c_{11}}{2}
({\bf \nabla}_\perp \cdot {\bf u})^2 
+ \frac{c_{66}}{2} ({\bf \nabla}_\perp \times {\bf u})^2 + 
\frac{c_{44}}{2} (\partial_z {\bf u})^2 \large].
\label{eq1}
\end{equation}
Here ${\bf u}({\bf r}_\perp,z)$ is the displacement field at location
(${\bf r}_\perp,z$). This term represents the elastic cost of
distortions from the crystalline state, governed by the values of the
elastic constants for shear (c$_{66}$), bulk (c$_{11}$) and tilt
(c$_{44}$).

Quenched random pinning destabilizes the long-ranged translational
order of the flux-lattice phase\cite{review1,larkin}. 
Such disorder adds 
\begin{equation}
F_{d} = \int d{\bf r}_\perp dz~V_d({\bf r_\perp},z)\rho({\bf r}_\perp,z),
\label{eq2}
\end{equation} 
to F$_{el}$ above, where F$_d$ represents the interaction of 
vortex lines with a quenched disorder potential $V_d$ and the  
density field $\rho({\bf r}_\perp,z)$ is defined by
\begin{equation}
\rho({\bf r}_\perp,z) = \sum_i \delta^{(2)} ({\bf r}_\perp - {\bf R}_i -
{\bf u}({\bf R}_i,z)).
\end{equation}

If the disorder derives from a high density of weak, random point
pinning sites, it is convenient to assume that $V_d$ is drawn from a
Gaussian distribution and is correlated over a length scale $\xi$, the
coherence length {\it i.e.} $[V_d(x)V_d(x^\prime)] = K(x - x^\prime)$,
with $K(x-x^\prime)$ a short-ranged function of range
$\xi$\cite{larkin}. The notation $x$ denotes (${\bf r_\perp},z$).

Much theoretical  attention has been devoted recently to the understanding of
the statistical mechanics defined by Eqns.(\ref{eq1}) and
(\ref{eq2}). The problem is the computation of correlation
functions such as
\begin{equation}
B({\bf r_\perp},z) = [\langle {\bf u}({\bf r}_\perp,z) - {\bf u}({\bf 0},0) 
\rangle^2],
\label{eq0}
\end{equation}
where the brackets $\langle\cdot\rangle$ and $[\cdot]$
denote thermal and disorder averages respectively. 
At least two novel glassy phases
arise as a consequence of such disorder\cite{fifi,natterman,giam}. In 
the relatively more ordered of these
phases, the Bragg glass (BrG) phase,
correlations of the order parameter for translational correlations
decay as power laws
\cite{natterman,giam,giam1,gingras,natterman1,emig}:
\begin{equation}
C_{\bf G}({\bf x_\perp}) = \overline{\left<\exp{i[{\bf G}\cdot 
({\bf u}({\bf x}_\perp,z) - {\bf u}({\bf 0},z))]}\right>}\sim 
1/|{\bf x}_\perp|^{\bar{\eta}_G},
\end{equation}
where ${\bf G}$ is a reciprocal lattice vector and 
$\bar{\eta}_G$ a non-universal
exponent \cite{natterman1,emig}.  The anomalously large translational
correlation lengths inferred from magnetic decoration experiments on
BSCCO\cite{decoration} and the resolution limited Bragg peaks obtained
in neutron scattering from the mixed phase\cite{neutron} in this
material at small $H$, support the existence of a Bragg glass phase in
type-II superconductors.  Hall probe measurements\cite{zeldov} see a
single first-order, temperature-driven melting transition out of the
Bragg glass at such fields\cite{neutron,zeldov,gammel,young}.

At increased levels of pinning, the Bragg glass is believed to be
unstable to a disordered phase in which translational correlations
decay exponentially at sufficiently large
distances\cite{giam1,gingras} {\i.e.}
$\lim_{x \rightarrow \infty} C_{\bf G}({\bf x}) \sim 
\exp(-\mu|{\bf x}_\perp|)$.
Such a phase {\em must} be separated by a
phase transition from the Brass glass phase.  At large $H$,
experiments\cite{experiments} indicate a continuous transition from an
equilibrium fluid phase into a highly disordered glassy state when the
temperature $T$ is reduced from the mean-field
$T_c(H)$\cite{usreview,vinokur}.  This disordered low-$T$, large-$H$
state may be a new thermodynamic phase, the ``vortex glass'' (VG)
phase, separated from the equilibrium disordered liquid (DL) by a line
of continuous phase transitions\cite{fifi}.  However, the experiments
may also be indicating a non-equilibrium regime\cite{frozen}.  A
``frozen liquid'' similar to a structural glass and an ``entangled
liquid'' analogous to a polymeric glass have been suggested as
alternatives to the equilibrium vortex glass\cite{frozen,ertaz}.

The phase diagram of Fig. 1 summarizes the currently popular view
of the phase behaviour in the mixed phase of weakly disordered
single crystals of anisotropic high-T$_c$ systems such as
BSCCO\cite{giam1,gammel,young,vinokur,kierfeld,kiervin,footnotedis}.
(This disorder is assumed to arise from a large density of weak point
pinning sites in this paper\cite{typedis}.)  Fig. 1 implies that the BrG
phase is unstable to the VG phase on increasing $H$. This feature follows
from the belief that increasing $H$ at a fixed level of microscopic
disorder is equivalent to increasing the {\em effective} level of
disorder\cite{gimcdg}.  Fig.  1 also shows the Bragg glass melting
{\em directly} into the DL phase in the intermediate field regime
(solid line).  The VG to DL transition begins where the first-order
BrG-DL transition line terminates; this transition has been suggested to
be continuous\cite{young}.  Recent Josephson plasma resonance experiments
and magneto-optic studies in BSCCO\cite{gaifullin,vanderbeek} indicate
that the underlying field-driven BrG-VG transition may actually be a
discontinuous one\cite{nonomura}.

The importance of thermal fluctuations in the cuprates ensures that
disordered phases such as the DL phase occupy much of the phase
diagram in high-T$_c$ materials\cite{review1}.  In contrast, for
most low-T$_c$ systems, such disordered phases occupy a relatively
small regime just below T$_c(H)$.  The phase behaviour in the mixed
phase of low-T$_c$ superconductors has been the focus of recent
attention\cite{satya1,satya2,satya3,jsps,mypaper}.  These studies
suggest a universal phase diagram for weakly pinned low-T$_c$ type-II
materials\cite{mypaper}. Interestingly, this phase diagram appears to
differ significantly from Fig.  1.

The low-T$_c$ systems studied have relatively large values of the
Ginzburg number,
\begin{equation}
G_i = (k_BT_c/H_c^2\epsilon\xi^3)^2/2,
\end{equation} which
measures the relative importance of thermal
fluctuations\cite{review1}. Here $\xi$ is the coherence length,
$\epsilon$ the mass anisotropy, $T_c$ the superconducting transition
temperature and $H_c$ the thermodynamic critical field. In most
low-T$_c$ materials $G_i \sim 10^{-8}$, whereas $G_i \sim 10^{-2}$ for
high-$T_c$ superconductors such as YBa$_2$Cu$_3$O$_{7-\delta}$ (YBCO)
and Bi$_2$Sr$_2$CaCu$_2$O$_{8+\delta}$ (BSCCO)\cite{review1}.

For the layered dichalcogenide 2H-NbSe$_2$ (T$_c \simeq 7.2 $K),
anisotropy, weak layering, small coherence lengths and large
penetration depths enhance the effects of thermal fluctuations,
yielding G$_i \sim 10^{-4}$\cite{shobo,shobo1}.  For the C15 Laves-phase
superconductor CeRu$_2$ (T$_c \simeq 6.1 $K), G$_i \sim
10^{-5}$\cite{chaddah}, while for the ternary rare-earth stannides
Ca$_3$Rh$_4$Sn$_{13}$, (T$_c \simeq 8$K) and Yb$_3$Rh$_4$Sn$_{13}$,
(T$_c \simeq 7.6$K) , G$_i \sim 10^{-7}$\cite{tomy1,tomy2,shampa}.  For the
quarternary borocarbide YNi$_2$B$_2$C (T$_c \simeq 15.5$K), G$_i \sim
10^{-6}$\cite{boro1}.  These systems can be made relatively pure: the
ratio of the transport critical current $j_c$ to the depairing current
$j_{dp}$ ($j_{dp} = cH_c/3\sqrt{6}\pi\lambda$, with H$_c = 
\Phi_0/2\sqrt{2}\pi\lambda\xi$) in these low-T$_c$ materials can be 
as small as $10^{-4}$ or $10^{-5}$. For the high-T$_c$ case, in 
contrast, $j_c/j_{dp} \sim 10^{-2}$.
These values of G$_i$ (1-4 orders of magnitude larger than for
conventional low-T$_c$ materials) and of $j_c/j_{dp}$ imply that the
phase behaviour of weakly pinned, low-T$_c$ type-II superconductors can
be studied using such compounds, in regimes where the phases and phase
transitions discussed in the context of the cuprates are experimentally
accessible.  

Fig. 2 presents the proposal of this paper for a
schematic phase diagram for type-II superconductors
with point pinning disorder.  This phase diagram
incorporates and extends recent suggestions for the
phase behaviour of low-T$_c$ systems, proposed by this
author and collaborators in Ref. \cite{mypaper}. It is
argued that Fig. 2 should be a universal phase diagram
for {\em all} type-II superconductors with quenched
point pinning disorder, both low and high-T$_c$, in a
sense made precise in this paper.

In Fig. 2, the term ``multi-domain glass'' (MG)
describes the sliver of intermediate glassy phase,
which we propose should generically intervene between
BrG and DL phases.  This phase is argued to be an {\em
equilibrium} phase. At present, neither experiments
nor simulations provide a clear indication of the
nature of this phase nor constrain the various
proposals made in this regard.  One possibility is
that the symmetry breaking which occurs across the
MG-DL transition line is related to the replica
symmetry breaking proposed recently in the context of
the structural glass transition\cite{parisi}. A
hexatic glass is an alternative which is attractive in
many respects\cite{hexatic}.

Modern replica-based approaches\cite{parisi} to the
glass transition in the context of fluids {\em
without} quenched disorder, which argue for a one-step
replica symmetry breaking transition out of the
disordered liquid phase, have recently been
generalized to study the phase diagram of hard spheres
in the presence of a quenched random
potential\cite{thalmann}.  Such a phase diagram bears
a fairly close resemblance to phenomenological phase
diagrams of disordered vortex systems, including the
existence of both first-order and continuous
transitions out of the disordered
liquid\cite{thalmann}.  The nature of symmetry
breaking here is subtle and it is unclear whether
experiments would be able to distinguish between a
genuine equilibrium glass transition of this type and
a non-equilibrium ``freezing-out'' of structural
degrees of freedom.  Nevertheless, in principle,
signals of such an equilibrium transition should be
visible in {\em thermodynamic} measurements.  Should
these ideas be applicable to the disordered mixed
phase, a {\em discontinuous} freezing of the
disordered density configuration characterizing the
glassy phase would be expected, as against relatively
{\em smooth} behaviour of the entropy and internal
energy across this transition.

It is believed that the lower critical dimension for
phase coherence of the type assumed in models for
vortex glasses exceeds three\cite{nattermann_private};
this would rule out the phase-coherent vortex
glass\cite{fifi} as an alternative.  However, the
absence of such long-ranged superconducting phase
coherence does not rule out the possibility of various
levels of order and associated symmetry breaking
transitions within the vortex line assembly.  It is to
be emphasized that the precise pattern of this
symmetry breaking is irrelevant to the phenomenology
embodied in Fig. 2, in particular to the proposal of a
generic two-step transition out of the BrG phase which
occurs via an intermediate equilibrium phase with
glassy attributes, in which length scales for
translational correlations in the flux line array can
far exceed those in the equilibrium fluid. Even if the
true ground state were an equilibrium vortex glass in
the sense of the original proposal, the topology of
the phase diagram proposed in this paper and most of
its attendent conclusions (with the specific exception
of those which relate to the superconducting
properties of this state) would remain
valid\cite{clarifying}.

The multi-domain glass at intermediate $H$ connects
smoothly to the highly disordered glassy state
obtained at large $H$, the putative vortex glass.  The
term ``multi-domain'' focuses attention on the
intermediate level of translational correlations
obtained within this phase in the interaction
dominated regime.  Melting on $T$ scans appears to
occur discontinuously out of the MG at these field
values\cite{lopez,contrast}.  Our MG phase is similar
to the recently proposed ``amorphous vortex
glass''\cite{kierfeld,kierfeld1}. Like the
``phase-incoherent vortex glass'' studied in detail in
Ref.\cite{natterman1}, our MG phase is topologically
disordered at the largest length scales\cite{vgnote}.
However, we emphasize that spatial correlations in the
MG phase can be fairly long-ranged, unlike
correlations in typical amorphous phases and in the
disordered liquid; also our MG phase is a genuine
phase, separated from the disordered liquid everywhere
by a line of phase transitions.  In addition, the
phase diagram we propose differs significantly from
ones proposed in earlier work.

The multi-domain character of this phase is argued to
manifest itself in the fact that the transition
between BrG and DL phases can occur via several
intermediate stages, corresponding to the melting of
different domains as a consequence of disorder-induced
inhomogeneities in the melting
temperature\cite{soibel}.  This can result in large
noise signals\cite{shobo} associated with a small
number of fluctuators\cite{shobo,merithew},
intermediate structure in the ac susceptibility as a
function of $T$\cite{satya1,jsps}, a stepwise
expulsion of vortices\cite{ooi1}, strong thermal
instabilities\cite{shobo,private} and a host of other
phenomena specific to this phase\cite{mypaper}, all
features of the experimental data, which are hard to
rationalize in other pictures.

We argue that the MG phase melts via a first-order
phase transition to the DL phase on $T$ scans at
intermediate $H$.  Fig. 2 shows this first-order
transition line terminating\cite{kiervin} at a
tricritical point, where it meets a (continuous) glass
transition line. (This topology is suggested by many
experiments\cite{young}; however, recent experimental
work, discussed in a later section, suggests a more
complex topology, involving a critical end point, for
the MG-DL transition line.) Some features of the phase
diagrams proposed in Ref.\cite{mypaper} and in Fig. 2,
notably the two-step character of the transition out
of the BrG phase in some regions of parameter space
and the reentrant nature of the phase boundaries at
low fields, were described in earlier
work\cite{satya1,satya2,satya3,jsps}.

This paper addresses the following question:  What
links the phase behaviour of low and high-T$_c$
systems?  Physically, it is reasonable to expect that
high-T$_c$ and low-T$_c$ superconductors should have
{\em qualitatively} similar phase behaviour. (This
expectation motivates some of the features of Fig. 2.)
However, Figs.  1 and 2 have some qualitatively {\em
different} features:  Fig. 2 implies that the Bragg
glass always melts directly into a disordered glassy
phase upon heating. This glassy phase is a
continuation of the disordered phase obtained on
cooling at high-field values. A subsequent transition,
at intermediate $H$, separates the MG from the
disordered liquid, as in Fig.  2.  In contrast,
consensus phase diagrams for high-T$_c$ systems such
as BSCCO and YBCO envisage at least one
``multicritical point'' (labelled (H$_{cr}$,T$_{cr}$)
in Fig.  1) on the BrG boundary in the $H-T$ plane.
Below this point the BrG melts directly into the
disordered liquid
\cite{giam1,natterman1,gammel,young,vinokur}.

This apparent distinction between the phase behaviour
of low and high-T$_c$ materials is surprising. This
distinction is also manifest in the difference in
phase behaviour exhibited by relatively pure samples
of high-T$_c$ materials which show an apparent
multicritical point as discussed above, and somewhat
more disordered samples.  Even among the low-T$_c$
systems, the data on relatively clean samples do not
rule out a possible multicritical point, in contrast
to the case for more disordered samples\cite{jsps}.

Thus, at least two possibilities suggest themselves.
We refer to these as scenarios (1) and (2). In the
first scenario, a genuine distinction exists between
more disordered and less disordered samples, leading
to the following picture for the experiments:  If the
disorder is sufficiently weak the sliver of MG phase
vanishes altogether and the BrG melts directly into
the DL phase\cite{some}. If the existence of a {\em
low}-field glassy phase is assumed (as shown in Fig.
2), there should be two multicritical points on the
BrG-MG transition line, where the high-field and
low-field glass transition lines meet the first-order
BrG-DL phase boundary.  As disorder is increased,
these two points approach each other on the BrG-DL
transition line. They then merge at a critical level
of disorder.  For stronger disorder, the BrG-MG
transition line splits off from the MG-DL transition
line, the two glassy phases are linked smoothly and
the phase diagram of Fig.  2 is obtained.

This scenario as well as the one outlined below {\em
assume} the existence of a low-field glassy phase just
above the lower critical field H$_{c1}$. Evidence for
this phase is strongest in the low-T$_c$ materials, in
particular 2H-NbSe$_2$,\cite{ghosh} although there
have been recent reports of similiar low-field
glassiness in the high-T$_c$ material
YBCO\cite{grover,hucho}.  The possibility of a
low-field glassy phase has been raised
theoretically\cite{nld}; simulations provide some
support to this hypothesis\cite{magro}. Recent
experiments indicate that the domain of this low-field
glassy phase can extend far above H$_{c1}$ in
disordered samples of 2H-NbSe$_2$\cite{jsps,paltiel0}.
In addition, the transition between the low-field
glassy phase and the BrG phase appears to be
sharp\cite{paltiel0}.

The second scenario is the following: Any distinction between samples
with different levels of disorder is only notional and the generic
phase behaviour is that shown in Fig. 2\cite{menon1}. Thus, the BrG
phase would always be expected to transform first into a disordered MG
phase and only then into the DL phase on $T$ scans.  It would then be
necessary to rationalize the experimental observations of a possible
single transition at intermediate $H$.  {\it A-priori}, both scenarios
are acceptable.  However, they cannot both be correct {\em if} the
phase diagram of a type-II superconductor with weak point pinning is
universal.

In this paper, it is suggested that the second scenario is a
theoretically attractive and experimentally viable alternative to the
first.  This paper discusses links between the phase diagrams of Figs.
1 and 2 in the context of the available experimental data, simulation
results and theoretical understanding. The phase
diagram of Fig. 2 is proposed to be the more basic one; Fig. 1 is
argued to be recovered from Fig. 2 when the BrG-MG and MG-DL transition
lines shown in Fig. 2 come so close as to be separately unresolvable.
(An alternative phase diagram which retains this feature but exhibits a
more complex topology for the MG-DL transition line is discussed in a
subsequent section.) It is argued that this is the case when thermal
averaging over disorder is substantial, leading to an effective
smoothening of the disorder potential. This discussion is motivated by
the following conjecture for phase transitions out of the BrG phase:
{\em The BrG phase is conjectured to always transform directly into the
MG phase and never in a single step into the DL phase, at any finite
level of disorder}\cite{menon1}.  Evidence from simulations, experiment
and theory which supports this conjecture is discussed.

This paper contains six further sections. Section II summarizes the
proposals of Ref.\cite{mypaper} for low-T$_c$ systems and then
supplements them with proposals appropriate for the description of
high-T$_c$ materials. Quantitative estimates are then provided in
support of these ideas.  Sections III and IV describe a theoretical
framework within which these proposals, particularly with regard to the
phase boundaries, can be quantified -- Section III discusses a Lindemann
parameter-based approach to obtaining the phase boundary which
separates BrG from MG phases while Section IV discusses a semi-analytic
approach to the MG-DL transition based on a replicated liquid state and
density functional theory.  Section V compares the predictions of these
sections with experimental data, principally on low-T$_c$ systems.
Section VI discusses the proposals of Section II in the context of the
experimental data on both high and low-T$_c$ systems.  The concluding
section (Section VII) links these results with previous work and
suggests several experiments which could explore the ideas described
here.

\section{Universal phase diagram for\\ 
disordered Type-II superconductors}

The proposals of Ref.\cite{mypaper} derived from an analysis of peak
effect systematics in the low-T$_c$ materials 2H-NbSe$_2$, CeRu$_2$,
Ca$_3$Rh$_4$Sn$_{13}$, Yb$_3$Rh$_4$Sn$_{13}$ and YNi$_2$B$_2$C.
Ref.\cite{mypaper}, extending earlier work\cite{jsps}, pointed out that
these systematics suggested a general link to an underlying {\em
static} phase diagram.  The peak effect (PE) itself refers to the
anomalous {\em increase} in $j_c$ seen close to $H_{c2}(T)$; this
increase terminates in a peak, as $H$ (or $T$) is increased, before
$j_c$ collapses rapidly in the vicinity of 
H$_{c2}$\cite{belincourt,campbell,kestsuei}.

The results of Ref.\cite{mypaper} relating to the
low-T$_c$ materials studied there are summarized in
the first three proposals listed below; the reader is
referred to the original paper for more details.
These proposals are extended to high-T$_c$ systems in
this paper. The approach is primarily phenomenological
-- it draws on published experimental data,
simulations and theoretical input in attempting to
describe a complete framework in which a large body of
data on phase behaviour in the mixed phase with
quenched point pinning disorder can be systematized
and understood.

\subsection{Proposals}
\newcounter{publist1}
\begin{list}{\arabic{publist1}.}{\usecounter{publist1}}

\item The high-field ``vortex glass'' survives at intermediate field
values as a sliver of disordered, glassy phase intervening between
quasi-ordered (BrG) and disordered liquid (DL) phases\cite{mypaper}. (We
use the phrase ``multi-domain glass'' (MG) to describe both the
intermediate sliver phase and the putative vortex glass phase to which
it connects smoothly.)  Thus, in going from Bragg glass to liquid on
increasing the temperature at intermediate field values, two phase
boundaries are generically
encountered\cite{satya1,satya2,satya3,ling0}. The first separates the
BrG phase from the MG phase while the second separates the MG phase from
the DL phase\cite{satya1,satya2,satya3,mypaper}.  At low field values,
the BrG-MG phase boundary is reentrant
\cite{nelson,ghosh,nld,magro,satya4}.  For materials with intermediate
to high levels of disorder, the glass obtained at high field values is
smoothly connected to the one obtained at low fields\cite{mypaper}.  At
low levels of disorder, the width of the sliver regime at intermediate
fields is extremely narrow\cite{satya1,jsps}.

\item The sliver of glassy phase at intermediate field values defines
the peak regime\cite{shobo}, the regime seen between the onset of the
increase of the critical current $j_c$ (at T$_{pl}$ on temperature
scans) and its peak value (at T$_p$ on temperature scans) as functions
of the applied field $H$ and temperature $T$ in materials which exhibit
a sharp peak effect\cite{jsps,mypaper}.  The dynamical
anomalies seen in experiments in this regime (slow dynamics, history
dependence, metastability, switching phenomena etc.) arise as a
consequence of the static complexity of the underlying glassy
state\cite{mypaper}.  Linking these anomalies to an underlying
equilibrium transformation of the flux-line system into a glassy state
({\it i.e.} the statics as opposed to the dynamics) suggests a simple
and general solution of the problem of the origin of PE anomalies.

\item Structure in the sliver phase obtained at intermediate fields
resembles a ``multi-domain'' structure, essentially an arrangement of
locally crystalline domains\cite{mypaper}.  The
possibility of such an intermediate state is attractive because it
represents a compromise between the relatively ordered lattice state
and the fully disordered fluid.  At the structural level, it is a
``Larkin domain solid'', originally believed to represent the fate of a
crystal on the addition of quenched
disorder\cite{larkinov1}.  This proposal rationalizes a large
body of data on PE anomalies including ``fracturing'' of the flux-line
array\cite{satya3}, the association of thermodynamic melting with the
transition into the disordered liquid\cite{lopez} as well as
conjectures regarding the dynamic coexistence of ordered and disordered
phases in transport measurements in the PE regime\cite{paltiel}.  

\item The transverse size of a typical domain $R_D$ in such a
``multidomain'' structure can be much larger than the
mean-interparticle spacing $a$. It is conjectured that R$_D$ is bounded
above by $R_a$, the length-scale at which thermal and disorder induced
fluctuations of the displacement field (calculated within the BrG phase)
are of order $a$\cite{kiervin}.  R$_D$ is also conjectured to be {\em
largest} in the intermediate-field, interaction-dominated
regime but should decrease rapidly for much larger
or smaller $H$, reflecting the increased importance of disorder both at
high-field and low-field ends\cite{jsps,mypaper,ghosh}. Similar
statements should hold for the longitudinal sizes of such domains.  In
such a phase, translational correlations would be solid-like out to a
distance of order R$_D$, but would decay rapidly thereafter. For
sufficiently weak disorder, the size of a typical domain is
large, reflecting the value of R$_a$.  In
contrast, in the DL phase significant spatial correlations exist only
only up to a few (1-2) intervortex spacings. Thus, the DL and MG phases
at intermediate $H$ should differ substantially at the
level of local correlations for sufficiently weak disorder.

\item At intermediate values of $H$, the glassy phase to which the
Bragg glass melts is confined to a narrow sliver which broadens out
both at much larger and much smaller $H$.  Thermal smoothening of
disorder reduces the width of this sliver. It can render the sliver
unobservable if the disorder is sufficiently weak, yielding an apparent
single melting transition out of the ordered phase. This can happen for
two reasons: First, domain sizes can be comparable to typical sample
dimensions if the effective disorder is weak enough.  Second, changes
in critical currents signalling a transition between different phases
can be too small to be resolvable; estimates also suggest that
magnetization discontinuities across the BrG-MG transition are very
small.  The apparent vanishing of the sliver explains the putative
multicritical point\cite{gammel,vinokur} invoked in the context of
phase diagrams for the mixed phase. While the sliver need not be
resolvable in some classes of experiments which probe phase behaviour
(for instance, it may not show up in dc magnetization measurements), it
may be apparent in other experiments, particularly those which probe
local correlations.  For more disordered superconductors,
whether of the high-T$_c$ or low-T$_c$ variety, a two-step transition
should generically be seen.

\item A vortex glass phase of the type proposed by Fisher, Fisher and
Huse may not exist. Gauge glass models for the VG transition do not show
a finite $T$ transition in three dimensions, once effects of screening
are included\cite{bokil}.  Also, recent work questions earlier reports
supporting a scaling description of the experimental data on disordered
cuprates \cite{strachan,brown}.  Theoretically, the evidence appears to
favour a low-temperature glassy phase {\em without} long-ranged phase
correlations, at least in the weak disorder limit\cite{natterman1}.
It is thus unlikely that the glassy phase obtained at low temperatures
and elevated field values is a true phase-coherent vortex glass,
in the sense of the original proposal\cite{natterman1}.  However,
many experiments in the peak effect regime of low and high-T$_c$
materials show that the multi-domain glass which intervenes
between the BrG and the DL phases has strongly glassy properties
\cite{satya1,satya2,shobo,merithew,henderson,marley,henderson1}.  The data
on low-T$_c$ systems favours an equilibrium transition\cite{mypaper}
for the following reasons:  (i) the approach to the intermediate regime
is discontinuous and the phase boundaries reproducible, (ii) behaviour in
this regime is dynamically very anomalous and, (iii) this phase connects
smoothly to the high-field (VG) phase, argued extensively to be glassy
in nature.  This suggests that the intermediate multi-domain glass is
a true equilibrium glassy phase for reasons which may have nothing to
do with the original vortex glass proposal.  The precise nature of the
symmetry breaking which distinguishes the MG phase from the DL phase is
at present unclear; a ``hexatic glass'' of vortex lines and a 
``replica symmetry broken glassy phase'' in the sense of modern
approaches to the structural glass transition are both attractive
possibilities.

\item As a consequence of such glassiness, a hierarchy of long-lived
metastable states can be obtained over significant parts of the phase
diagram, leading to slow and non-trivial relaxation
behaviour\cite{henderson,satya6}.  Such states may dominate the
experimentally observed behaviour at low temperatures and high fields
and mimic the behaviour expected of an equilibrium vortex glass phase.
As a further corollary, experiments at low temperature and high field
values may often be probing strongly non-equilibrium behaviour, once
time-scales for structural relaxation exceed experimental time-scales,
thereby obscuring the underlying {\em equilibrium} phase transitions.

\end{list}

\subsection{Evidence for a Multi-domain state in the MG phase}

A substantial body of data favours a multi-domain structure in the
regime intermediate between BrG and DL phases. A representative
cross-section of this evidence is presented here.

Recent neutron scattering measurements on disordered single crystals of
Niobium (H$_{c2}(4.2K)=4.23~kOe$,T$_c(0)\simeq 9K$), which show a sharp
peak effect, suggest fairly ordered crystal-like states obtained by
heating following cooling in zero field, even for temperatures in the
regime between T$_{pl}$ (T$_o$ in Ref.\cite{ling}) and T$_p$, the
``peak regime''\cite{ling}. Following Ref.\cite{mypaper}, 
this regime defines the sliver of MG phase
which intervenes between BrG and DL phases at intermediate $H$.
In contrast, the metastable disordered states obtained
by field-cooling through the peak regime show nearly isotropic rings of
scattering, indicating relatively short translational correlation
lengths, comparable to those in a disordered liquid state.  Experiments
see a 30\% reduction in rocking curve widths of Bragg spots between
field cooled (FC) and zero-field cooled (ZFC) configurations
demonstrating a fair degree of local order in the latter.  That the
relatively ordered states in the peak regime obtained on zero field
cooling are actually lower in free energy can be demonstrated through
experiments in which field-cooled states below T$_p$ are annealed
through the application of an ac field\cite{ling}.  These data are
consistent with SANS data on 2H-NbSe$_2$\cite{private}.

Recent muon-spin rotation experiments\cite{blasius,review_muon} 
on BSCCO show a narrow intermediate
phase where the asymmetry parameter associated with the field
distribution function $n(B)$,
\begin{equation} 
\alpha = \langle\Delta B^3\rangle^{1/3}/\langle\Delta B^2\rangle^{1/2},
\end{equation} 
where $\Delta B$ = $B - \langle B \rangle$, jumps from a value of
$\alpha \sim 1.2$ at low $T$ to a somewhat smaller value $\alpha \sim
1$ before a further jump to a value close to zero\cite{blasius} at the
irreversibility line. Large values of this asymmetry $(\alpha \sim
1.2$) indicate a fairly well-ordered lattice structure, while values
close to zero indicate a highly symmetric arrangement of vortices, as
in a liquid.  The observation of such asymmetric linewidths in this
narrow intermediate phase can be associated to the multidomain
structure of the sliver of MG phase expected to intervene between BrG
and DL phases.  These results would suggest a structure with a fair
degree of local order, sufficiently different from the disordered
liquid.

As $T$ is raised further, individual domains can melt due to disorder
induced variations in the local melting temperature $T_M$, leading to a
mosaic of solid-like and liquid-like (amorphous) domains, similar to a
regime of two-phase coexistence.  This picture rationalizes several
observations in the literature.  Experiments on YBCO find that the
magnetization discontinuities and first-order behaviour associated with
the melting transition in the interaction dominated regime at
intermediate $H$ can be associated with the {\em MG-DL} transition at
higher fields, once the BrG-MG line splits off from the putative
multicritical point\cite{ssten,nishizaki}.  In addition, there appears
to be a critical point on this first-order line, beyond which these
discontinuities vanish.  The idea that structure similiar to two-phase
coexistence may account for some properties of the vortex array just
below the melting transition has been raised in earlier
work\cite{gordeev,rassau}.

These features are easily explained within the picture outlined here.
The melting transition in this scenario should generically be
associated with the MG-DL shown in Fig.  2, although the finite width
of the sliver regime implies that some part of the discontinuity in
density between fluid and solid phases can be absorbed across the width
of the sliver\cite{mypaper}.  If $R_D$, the characteristic domain size,
is of order or smaller than the translational correlation length at
freezing $\xi_M$, sharp signals of melting are no longer
expected\cite{gimcdg}.  On the other hand, if $R_D \gg \xi_M$ at the
melting transition, the behaviour should be equivalent to that of the
pure system. Assuming $R_D \gg \xi_M$ initially, as $R_D$ decreases on
an increase in $H$ (proposal (4) and the results of
Refs.\cite{kiervin,kierfeld1}) it approaches $\xi_M$; the point where
$R_D\simeq \xi_M$ defines the point where magnetization discontinuities
should cease to be seen and thus the experimentally observed critical
point.

The possibility that the vortex glass phase obtained from the BrG phase
on {\em field} scans might have a relatively low equilibrium density of
dislocations was proposed recently by Kierfeld and Vinokur to
rationalize some of the anomalies mentioned above\cite{kiervin};
related ideas in the low-T$_c$ context appeared in Ref.\cite{satya1}.
We argue here that such a picture is also applicable to the {\em
temperature}-driven transition out of the BrG phase, as is clear from
Fig. 2.  

Hypothesizing such a multi-domain in the intermediate sliver regime is
consistent with the suggestion that a ``fracturing transition'' from a
solid with quasi-long-range order into a intermediate state with
correlation lengths of some tens of interparticle spacings is
responsible for the unusual open hysteresis loops obtained on thermal
cycling within the peak effect regime in 2H-NbSe$_2$\cite{satya1}.  The
idea that the regime which interpolates between the low-temperature
ordered phase and the high-temperature disordered phase might have a
relatively large degree of translational order has been used to
rationalize the difference in behavior of the ac susceptibility in ZFC
and FC scans in studies of the peak effect in several low-T$_c$
materials\cite{satya3}.

A fourth piece of evidence favouring a multi-domain or ``fractured''
state comes from voltage noise measurements, as pointed out in
Ref.\cite{mypaper}. Experiments see a substantial enhancement in noise
within the peak regime\cite{merithew,marley}, a feature also seen in
the ac susceptibility noise\cite{satya3}. This noise is profoundly
non-Gaussian in a regime where the PE is strongest; the number of
independent fluctuators contributing to this noise turns out to be
small\cite{merithew}. Experiments indicate a possible static origin for
this noise, as opposed to a purely dynamic origin such as the
interaction of fluctuating flow channels\cite{merithew}. This supports
a picture of domains of mesoscopic size {\em in equilibrium} with each
domain an ``independent fluctuator'' contributing to the noise.

Recent simulation work on a pancake vortex model\cite{olsen}
indicates that the PE may be associated primarily with a {\em
decoupling} transition in the c-axis direction, in which c-axis
correlations of vortices become short-ranged.  This author and
collaborators have argued recently that modelling the MG phase in a
pancake vortex model in terms of a random stacking of perfectly
crystalline layers should yield a fairly high-energy situation, with an
energy which scales linearly with the area of each plane\cite{frozen}.
It is reasonable that some fraction of this cost can be relaxed by
allowing dislocations to enter into each layer, leading to a situation
far more like the ``multidomain'' structure proposed in this paper than
a decoupled pancake-vortex state\cite{frozen,usreview}.

Other simulations, such as those of van Otterlo {\it et.
al.}\cite{otterlo1,otterlo2} are consistent with an intermediate field
MG phase with a correlation length intermediate between the solid and
the disordered liquid. These simulations see primary peaks in the
structure factor appropriate to the MG phase which transform into rings
only across a second transition into the DL phase. The existence of
these peaks indicates a fair degree of translational order, comparable
to the sizes of the system studied, although the dynamics changes
abruptly across the BrG-MG transition.  Interestingly, van Otterlo {\it
et. al.} conclude that they cannot rule out the existence of a sliver
of MG phase always preempting a direct BrG-DL transition\cite{mypaper}.

\subsection{Domain Sizes in the MG phase}

The scale for R$_a$ is itself set by a combination of the Larkin
pinning length scale, at which disorder and thermal fluctuation induced
displacements are of order $\xi$, as well as the inter-line spacing $a$
\cite{review1,natterman1}. These can be derived from $B({\bf
r}_\perp,z)$.  The transverse Larkin pinning length $R_p$ is defined
through $B(R_p,0) \simeq \xi^2$.  A second length-scale, the
longitudinal Larkin pinning length $L^b_p$, is defined through
$B(0,L^b_p) \simeq \xi^2$.  At transverse (longitudinal) length scales
of order $R_p$ $(L^b_p)$, the disorder-induced wandering of the
displacement field equals the size of a pinning centre ($\xi$) at
$T=0$.  A third important (transverse) length scale is the scale $R_a$
at which disorder and thermal fluctuation induced displacements in the
direction transverse to the field become of order the inter-line
spacing $a$:  $B(R_a) \simeq a^2$.

At finite temperatures, replacing $\xi^2$ by max$(\xi^2,<u^2>)$ in the
equations defining $L_p$ and $R_p$ above defines length scales $L^b_c$
and $R_c$. Here $<u^2>$ is the square of a ``Lindemann length''; these
quantities are directly related to the critical current.  In the Bragg
glass, $B(r)$ shows the following properties:  For $r_\perp,z \ll
R_c,L^b_c$, correlations behave as in the Larkin ``random force'' model
{\it i.e.} $B(x) \sim x^{4-d}$. At length scales between $R_c$ and
$R_a$ (the scale at which disorder-induced positional fluctuations
become of order $a$), the random manifold
regime, $B(x) \sim x^{2\zeta_{rm}}$. The exponent $\zeta_{rm} \sim
(4-d)/6 \sim 1/6$.  At still larger length scales, $B(x) \sim$ log(x),
yielding an asymptotic power law decay of translational
correlations\cite{natterman,giam,giam1,natterman1}.

The Larkin length scales $R_p$ and $L^b_c$ and the
Larkin volume $V_c \sim R_c^2L^b_c$ are estimated as\cite{review1}
$L^b_c = 2\xi^2c_{66}c_{44}/nf^2$
and
$R_c = \sqrt{2}\xi^2c^{3/2}_{66}c^{1/2}_{44}/nf^2$.
Using these, the standard weak-pinning expression for
the critical current density $j_c$ follows\cite{review1}:
\begin{equation}
j_c = \frac{1}{B}f\left(\frac{n}{V_c}\right)^{1/2}.
\end{equation}

R$_c$ and $L^b_c$ for a relatively clean 2H-NbSe$_2$ single crystal are
estimated in the following way\cite{review1}. We anticipate that the
following condition is satisfied: $R_c, L^b_c > \lambda$,  implying
that non-dispersive ($k=0$) values of the elastic constants can be used
in our estimates. Then,
\begin{equation}
j_c \sim \left(\frac{\xi}{R_c}\right)^2j_{dp},
\end{equation}
\begin{equation}
L^b_c \sim \frac{\lambda}{a} R_c.
\end{equation}

Assuming values of T$_c \simeq 7 K$, $\lambda 
\simeq 700 \AA$ ($H \parallel c$), $\xi \sim 70\AA$
and a flux-line
spacing $a (= \sqrt{2\Phi_0/\sqrt{3}B}) \sim 450 \AA $ at
$B \sim 1T$. For $10^{-4} < j_c/j_{dp} < 10^{-6}$, we
then obtain:
\begin{equation}
\frac{R_c}{a}\sim \cases{150~(j_c/j_{dp} = 10^{-6}); \cr
                         15~~(j_c/j_{dp} = 10^{-4})\cr}
\end{equation}
and
\begin{equation}
\frac{L^b_c}{a}\sim \cases{240~(j_c/j_{dp} = 10^{-6}); \cr
                         24~~(j_c/j_{dp} = 10^{-4})\cr}
\end{equation}

The domain size $R_D$ has been conjectured above (see proposal (4))
to be of order $R_a$ in the 
intermediate $H$, interaction dominated regime.  R$_a$
is estimated via
\begin{equation}
R_a \simeq R_c(\frac{a}{\xi})^{\large {1/\zeta_{rm}}},
\end{equation}
where $R_c$ is the (transverse) Larkin pinning length
scale\cite{giam,giam1,natterman1}. 
At typical laboratory fields $\sim 1T$, we estimate 
\begin{equation}
(\frac{a}{\xi})^{1/\zeta_{rm}} \sim (\frac{450}{70})^6 \sim 7\times 10^4.
\end{equation}
This leads to 
\begin{equation}
10^5 \leq  R_a/a \leq 10^6,
\end{equation}
for a system with the parameter values described above.  These are
{\em overestimates}; allowing for the wave-vector dependence of elastic
constants and prefactors neglected in these simple order-of-magnitude
estimates should reduce these numbers considerably, possibly
by factors of 10 or more.

Assuming a multi-domain structure within the peak regime, a
typical scale for each domain of $R_a \sim 10^4$a, $a \sim 400 \AA$
at $H \sim 1T$
and a (longitudinal) Larkin length comparable to the size of the
sample in the c-axis direction, one can obtain an estimate for the
number of ``independent fluctuators'' discussed above in
the context of the noise measurements. For a crystal of
typical transverse area $A \sim 1 mm \times 1 mm 
\sim 10^{14}\AA^2$ the number of such 
fluctuators $N_f$ is estimated as
\begin{equation}
N_f =  \frac{A}{A_D} \sim  10^1-10^2,
\end{equation}
numbers small enough to yield the strong non-Gaussian
effects seen in the experiments, particularly if one assumes
that not all domains are active contributors to the
noise signal.

Thermal fluctuations smear the effective
disorder potential seen by a vortex line. If the thermally induced
fluctuation of a vortex line about its equilibrium position exceeds
$\xi$, the effective pinning
potential seen by a pinning site is strongly renormalized. An estimate
of this effect at the level of a single pinned line
\cite{nld} yields
\begin{equation} 
U_p \sim U_p(0)\exp[-c(T/T_{dp})^3], 
\end{equation} where U$_p(0)$ is the
pinning potential associated with a single pinning site at $T=0$, $c$
is a numerical constant and T$_{dp}$ is a characteristic temperature
scale, the depinning temperature. 

The value of the depinning temperature T$_{dp}$ can be estimated
from
\begin{equation}
T_{dp} \sim  (U_p^2\xi^3c\epsilon_0/\gamma^2)^{1/3}.
\end{equation}
$U_p$ is a measure of the pinning energy per unit length, $\epsilon_0$
is the line tension of a single vortex, $c$ is a numerical constant of
order unity and $\gamma$ is the mass anisotropy. Independent measures
yield T$_{dp} \sim 25-40K$ for YBCO\cite{ertaz,kierfeld,ssten}.

The ratio $R = T^{low}_{dp}/T_{dp}^{high}$ is then a quantitative measure
of the relative importance of thermal fluctuations in low and high
T$_c$ materials. We use the following values:
\begin{equation}
NbSe_2: U_p = 10K/\AA, \xi = 70\AA, \lambda = 700\AA,
\gamma = 5
\end{equation}
and 
\begin{equation}
YBCO: U_p = 10K/\AA, \xi = 20\AA, \lambda = 1400\AA, \gamma = 50,
\end{equation}
and obtain
$R \sim 10^2$, suggestive of the relative importance of this effect to
the high-T$_c$ materials {\it vis a vis} its irrelevance in low-T$_c$
systems except very close to H$_{c2}$. A more accurate estimate of
the magnitude of this effect comes from computing the ratio 
\begin{equation}
R\frac{T^{high}_m}{T^{low}_m} \sim R\frac{T^{high}_c}{T^{low}_c} \sim 10^3,
\end{equation}
where T$_m$ is the melting temperature of the pure flux-line lattice
and we have approximated T$_m \sim $T$_c$ in both cases.

That thermal fluctuations lead to a substantial reduction in the
effectiveness of disorder close to the melting line in YBCO and BSCCO
is clearly evident from experimental measurements of the
irreversibility line in single crystals of these
materials\cite{ssten,nishizaki,nishizaki1}. The irreversibility line
at intermediate fields ($H \sim 3-7T$) in YBCO actually
appears to lie well {\em below} the melting line in weakly disordered
samples\cite{ssten}. Thus, thermal melting occurs in the
almost totally reversible regime. Precision measurements\cite{billon}
reveal a weak residual irreversibility, with attendent j$_c$ values
close to the melting transition of about $0.4 A/cm^2$, comparable to
that obtained in the purest samples of 2H-NbSe$_2$, but at much larger
temperatures; T$_m \sim 80K$ in YBCO at $H \sim 5T$, compared to
T$_c$'s of about 7K for 2H-NbSe$_2$. Similiar results have been
obtained for BSCCO at fields less than about 0.05T\cite{majer}. In
contrast, experiments on the low-T$_c$ materials discussed in
Refs.\cite{jsps,mypaper} indicate an irreversibility line which lies
{\em above} both T$_p$ and T$_{pl}$ lines, and subdivides the
disordered liquid further into irreversible and irreversible liquid
regimes. In these systems, the evidence for a two-step transition
appears unambiguous.

Such anomalously {\em small} values of j$_c$ (j$_c$ is proportional to
the width of magnetic hysteresis loops and vanishes in the perfectly
reversible regime), should lead to anomalously {\em large} values of
$R_c$.  Using j$_c/j_0 \sim 10^{-7}$, and $\xi$ and $\lambda$ values
appropriate to YBCO, yields a characteristic domain size of order $1
mm$, clearly comparable to typical sample dimensions.  Although these
estimates are approximate ones, the physical intuition should be
robust: characteristic domain sizes associated with an intermediate MG
phase in high-T$_c$ materials can be fairly large.  Accordingly, in
relatively pure samples of a small size, only a single sharp melting
transition may be obtained, justifying the putative multicritical point
and the direct transition into the liquid phase obtained in some
measurements.

The estimates above indicate that the length $R_a$ can be far larger
than typical correlation lengths in the disordered fluid at the melting
transition $\xi_M$ $\sim  2-3 a$.  SANS experiments on Nb indicate
translational correlations of this order\cite{ling}.  Thus sharp
signals of melting can be expected at the second transition line
intersected on a $T$ scan out of the BrG phase; such signals should
cease when $R_D \sim \xi_M$.

The phase boundary between a multidomain solid and the disordered
liquid can thus be argued on physical grounds to have a critical
point.  This is analogous to the gas-liquid transition in the following
way: There is no symmetry distinction at the {\em structural} level
between a multi-domain state and a DL phase; the multi-domain solid has
neither long-ranged crystalline order nor the power-law correlations of
a Bragg glass.  However, if a thermodynamic transition into a glassy
phase with a symmetry different from that of the DL (or multi-domain
state) exists, the associated phase boundary cannot terminate except at
$T=0$ or {\em on} another phase transition line.  Fig.  2 accounts for
such an equilibrium glass transition. The topology of the MG-DL phase
boundary is consistent with recent experimental work on single crystals
of YBCO\cite{ssten} and data on low-T$_c$ systems\cite{jsps,mypaper}.
However, no arguments appear to rule out alternative locations for this
line of glass transitions such as the one proposed in recent simulation
work\cite{nonomura} or more complex topologies, such as a critical
end-point for the MG-DL transition line.

\section{BrG-MG Phase Boundary from a Lindemann Parameter Approach}

The field-driven BrG-MG transition occurs even at $T=0$, where it is
driven solely by changing the effective disorder. It represents a
transition into a multidomain state at intermediate $H$ on $T$ scans.
Such a state is substantively different from the equilibrium disordered
fluid in terms of its local correlations and average density, provided
the disorder is sufficiently weak.  In contrast, the second transition,
between MG and DL phases, is a close relative of thermal melting in the
pure system.

The simplest way to estimate phase boundaries such as those shown in
Figs. 1 and 2 uses a Lindemann-parameter based approach.  The treatment
of the BrG-MG transition line described here uses arguments similar to
those of Giamarchi and Le Doussal (GLD)\cite{giam1,young}. 
It differs from their approach in the following way: GLD study this
transition only at $T=0$ (where thermal fluctuations are absent) and at
high temperatures where it is argued that disorder can be neglected and
an expression for the melting line in the pure system used. The
calculation here studies the crossovers in detail in the context of the
experimental data on the low-T$_c$ systems analysed in
Ref.\cite{mypaper}. A more substantial difference between this approach
and that of GLD is the proposal here that the high-temperature
intermediate field transition out of the Bragg glass is {\em not}
equivalent to thermal melting but rather to a transition to an
intermediate `multi-domain state'' with the properties discussed in the
previous section.

A convenient phenomenological characterization of the melting
transition in simple three-dimensional solids indicates that the
transition occurs when the root mean-square fluctuation of an atom from
its equilibrium position equals a given fraction
($c_L$, the Lindemann parameter) of the interatomic spacing $a$.
Giamarchi and Le Doussal suggested a possible generalization of this
idea to the study of the instabilities of the Bragg glass phase\cite{giam1}.
Taking over their proposal, we compute the ratio 
\begin{equation}
B(r_\perp = a)/a^2 = c_L^2,
\end{equation}
(see Eq.\ref{eq0}), to obtain the melting line taking c$_L$ to be a 
universal number independent of $H$ and $T$.

In the random manifold regime, using a variational replica symmetry
breaking ansatz, GLD derive the following relation
\begin{equation}
B(r_\perp) = \frac{a^2}{\pi^2}{\tilde b}(r_\perp/R_a),
\end{equation}
where the crossover function 
\begin{equation}
\tilde{b}(x) \sim 1,~~~~~~~~~~~ x \rightarrow 1,
\end{equation} 
and 
\begin{equation}
\tilde{b}(x) \sim x^{1/3},~~~~~~~~~~ x \rightarrow 0.
\end{equation} 

In a calculation
which assumes wave-vector independent elastic constants,
GLD obtain $R_a$ as
\begin{equation}
R_a = \frac{2a^4 c_{66}^{3/2}c_{44}^{1/2}}{\pi^3\rho_0^3U_p^2(2\pi\xi^3)}.
\end{equation}
Here $c_{44}$ is the tilt modulus of the flux-line system and $c_{66}$
its shear modulus, both calculated at zero wave vector.  $U_p$ measures
the pinning strength per unit length, $a$ is the mean intervortex
spacing and $\rho_0$ is the areal density of vortices given by $\rho_0
= B/\Phi_0$, with $\Phi_0$ the flux quantum: $\Phi_0 = 2.07 \times
10^{-15} T m^2$. It is assumed that $H \gg H_{c1}$, so that the
effects of bulk distortions can be neglected.

These expressions can be used to derive the full BrG-MG transition
line in the $H$-$T$ plane. For weak disorder, $R_a \gg a$. The 
following expressions for the elastic constants are used:
\begin{equation}
c_{66} = \frac{\epsilon_0}{4a^2}(1-B/B_{c2}(T))^2,
\end{equation}
and 
\begin{equation}
c_{44} = \frac{B^2}{4\pi}\frac{(1-B/B_{c2}(T))}{B/B_{c2}(T)}.
\end{equation}
Here $\epsilon_0 = (\Phi_0/4\pi\lambda^2)$.
The effects of a wave-vector dependent elasticity have been
approximately accounted for in writing these expressions.  
Note that the elastic constants vanish as 
$B \rightarrow B_{c2}(T) $; at lower fields the expression for c$_{66}$ should
include a contribution from the line tension of isolated flux
lines, which is neglected here.  

Some algebra then yields the intermediate expression
\begin{equation}
c_L^2 =  \frac{1}{a^{2/3}}\lambda(T)\frac{[B/B_{c2}(T)]^{1/6}}
{[1-B/B_{c2}(T)]^{7/6}}
\frac{1}{\Phi_0^{4/3}}(U_p^{2/3}\xi) \times N,
\end{equation}
where $N$ is a numerical factor.

To simplify $U_p^{2/3}\xi$, we use the ``core pinning'' assumption
\begin{equation}
U_p = p \frac{H_c^2}{8\pi}\xi^2,
\end{equation}
where $p$ is a constant of order 1. This models the
sources of pinning disorder as small-scale point
defects on the scale of $\xi$, which
act to reduce T$_c$ locally. Using 
\begin{equation}
H_c = \Phi_0/(2\sqrt(2)\pi\xi\lambda),
\end{equation}
leads to 
\begin{equation}
U_p^{2/3}\xi = \frac{\Phi_0^{4/3}}{(2\sqrt(2)\pi)^{4/3}}
(\frac{p}{8\pi})^{2/3}\frac{1}{\kappa}\frac{1}{\lambda^{1/3}},
\end{equation}

Defining $b = B/B_{c2}(T=0)$ and assuming a temperature
independent $\kappa (=\lambda/\xi)$, considerable algebra then yields 
the implicit expressions for the transition line $b(t)$ given 
below, given the following assumptions about the temperature 
dependence of the penetration depth $\lambda$ and the upper 
critical field H$_{c2}$:

\begin{enumerate}
\item Assuming a temperature dependence of $\lambda$ of the
phenomenological two-fluid type {\it i.e.}
\begin{equation}
\lambda = \lambda(T=0)/(1-t^4)^{1/2},
\end{equation}
where $t = T/T_c(H=0)$ and a linear
dependence of $B_{c2}(T)$, {\it i.e.} 
\begin{equation}
B_{c2}(T) = B_{c2}(0)(1-t),
\end{equation}
yields the relation:
\begin{equation}
\Sigma = \frac{b^3(1-t)^6}{(1-t^4)^2(1-t-b)^7}.
\label{eqbgvg1}
\end{equation}

Here $\Sigma$ is assumed to be constant; its value
involves the product of the Lindemann parameter c$_L$,
$\kappa$, and $p$ in the following ratio: $c_L^{12}\kappa^2/p^4$,
multiplied by a numerical constant of value $1.88 \times 10^7$. 
Assuming $\kappa \sim 1$ 
c$_L \sim 0.22$ and $p \sim 1$, yields 
$\Sigma \sim 0.25$. However, $\Sigma$ is very sensitive
to the value of c$_L$ used; changing $c_L$ by a factor of
2 to 0.11 changes $\Sigma$ by a factor of about 
$5 \times 10^{-3}$. For $c_L = 0.15, \kappa = 1$ and
$p = 1$, we obtain $\Sigma = 0.002$.
For this reason, we use $\Sigma$ as a fitting 
parameter and consider values for it in the range 
$10^{-3} < \Sigma < 10^2$. 

Eq. (\ref{eqbgvg1}) is plotted in Fig. 3, for different values of
$\Sigma$, in the range $0.01 < \Sigma < 60$; the 
MG phase lies above these lines while the BrG phase lies
below them. Note that the transition lines are almost
linear in (1-t) for large values of $\Sigma$.

\item Assuming a temperature dependence of $\lambda$ of the
phenomenological two-fluid type {\it i.e.}
\begin{equation}
\lambda = \lambda(T=0)/(1-t^4)^{1/2},
\end{equation}
where $t = T/T_c(H=0)$ and a quadratic
dependence of $B_{c2}(T)$, {\it i.e.} 
\begin{equation}
B_{c2}(T) = B_{c2}(0)(1-t^2),
\end{equation}
yields:
\begin{equation}
\Sigma = \frac{b^3(1-t^2)^6}{(1-t^4)^2(1-t^2-b)^7}.
\label{eqbgvg2}
\end{equation}

Again, the numerical value of $\Sigma$ is dictated essentially 
by the value of $c_L$; we choose a similiar range of $\Sigma$
values as in the previous case.
This relation is plotted in Fig. 4, for different values of
$\Sigma$, in the range $0.01 < \Sigma < 50$. These plots show
a more substantial curvature than the analogous plots for
case (1) above.
\end{enumerate}

In addition, assuming a temperature dependence of $\lambda$ of the
following linear type $\lambda = \lambda(T=0)/(1-t)^{1/2}$ where $t =
T/T_c(H=0)$ and a linear dependence of $B_{c2}(T)$, {\it i.e.}
$B_{c2}(T) = B_{c2}(0)(1-t)$ yields $\Sigma^{\prime\prime} =
\frac{b^2(1-t)^5}{(1-t-b)^7}$. This parametrization yield a BrG-MG phase
boundary which is virtually a straight line, for the range of $\Sigma$
values displayed in Figs. 3 and 4.  Note that all these results use
different approximate parametrizations of $B_{c2}(T)$; the
parametrizations involved in Eqs.(\ref{eqbgvg1}) and (\ref{eqbgvg2}) are
both commonly found in the literature.  Whether one or the other
expression should be used should depend on the quality of fits to the
experimentally obtained $B_{c2}(T)$ line. We use the first
parametrization for the 2H-NbSe$_2$ data (Section V), but the second
parametrization for both the CeRu$_2$ and the Ca$_3$Rh$_4$Sn$_{13}$
data discussed in that section.

For the phase boundaries to be physical, $t+b$ must be
less than or equal to 1. We perform the computations
using Eq.(\ref{eqbgvg1}).
One point on the curve is $(b=0,t=1)$; $b$ increases monotonically
with $1-t$. At $t=0$, the critical 
value of the magnetic
field separating the BrG phase from the MG phase satisfies
$\frac{b(0)^3}{(1-b(0))^7} = \Sigma$.
For $\Sigma$ small $b(0)\sim \Sigma^{1/3}$, while for large $\Sigma$,
$b(0) \sim 1 - 1/\Sigma^{1/7}$
The shape of the $b-t$  phase diagram close to $b=0$ is also easily
obtained. For $t \rightarrow 1$, we have
$b \sim (1-t)$.

These calculations use a simple analytic parametrization of correlation
functions in the BrG phase, obtained after many further approximations
on an initially simplified Hamiltonian, together with the restriction
to a constant (field and temperature independent) Lindemann parameter.
As a consequence, the quantitative predictions of this section should
not be taken excessively seriously. However, the qualitative trends in
the data appear to be borne out in this scheme of calculation, as
discussed further in Section V.

\section{Semi-analytic Calculation of the MG-DL Transition line}

This section discusses a calculational approach to the MG-DL phase
boundary.  The method outlined here uses results from a replica theory
proposed by this author and Dasgupta (Ref.\cite{gimcdg}) for the
correlations of a vortex liquid in the presence of random point
pinning\cite{usreview}. This work examined the instability, within
mean-field theory, of the DL phase to a crystalline state.

The estimates obtained above indicated that typical domain sizes in the
MG phase, could be much larger than the translational correlation
length at freezing in systems with low levels of pinning. In addition,
the transition out of the DL phase at intermediate $H$ appears
experimentally to be first order.  A natural first approximation is to
analyse, in mean-field theory, the instability of the liquid to an
ordered crystalline state.  This approach should thus represent the
physics of the first-order part of the MG-DL transition line in the
interaction dominated regime at intermediate $H$.  The issue of the
nature of glassiness in this phase cannot be addressed by these
methods.  However, these calculations should provide a useful upper
bound on the location of the actual MG-DL transition line.

The calculations of Ref.\cite{gimcdg} applied to a model
for BSCCO which considered only the {\em electromagnetic} interactions
between pancake vortices moving on different layers, ignoring their
(far weaker) Josephson couplings\cite{ldon}. Such a model is {\em
exact} for a layered system in the limit of infinite anisotropy.  For
large but finite anisotropy, its predictions are quantitatively fairly
accurate.  It is argued here in some detail at the end of this section
that the generic features of the results are relevant for the far more
isotropic materials discussed in this paper.

The analysis of Ref.\cite{gimcdg} was based on
the replica method~\cite{edwan,nld} applied to a system of 
point particles interacting via the Hamiltonian
\begin{equation}
H = H_{kinetic} + \frac{1}{2}\sum_{i \ne j} V(\mid{\bf r}_i - {\bf r}_j\mid)
+ \sum V_d({\bf r}_i),
\label{eq38}
\end{equation}
where $V(r)$ is a two-body interaction potential between the particles
and $V_d({\bf r})$ is a quenched, random, one-body potential, drawn
from a Gaussian distribution of zero mean and short ranged
correlations: $[V_d({\bf r})V_d({\bf r^\prime})] = K(\mid{\bf r} - {\bf
r}^\prime\mid)$, with $[\cdots]$ denoting an average over the
disorder.

Using $[\ln Z] = \lim_{n \rightarrow 0} [(Z^n -1)/n]$, one obtains,
prior to taking the $n \rightarrow 0$ limit, a replicated
and disorder averaged configurational partition function of the form
\begin{equation}
Z^R = \frac{1}{(N!)^n}\int\Pi d{\bf r}_i^\alpha
\exp(-\frac{1}{2k_BT}\sum_{\alpha=1}^n\sum_{\beta=1}^n
\sum_{i = 1}^N\sum_{j=1}^N
V^{\alpha\beta}(\mid{\bf r}_i^\alpha - {\bf r}^\beta_j\mid)).
\label{eq39}
\end{equation}
Here $\alpha,\beta$ are replica indices  and
$V^{\alpha\beta}(\mid{\bf r}^\alpha_i - {\bf r}^\beta_j\mid)
=V(\mid{\bf r}^\alpha_i - {\bf r}^\beta_j\mid)\delta_{\alpha\beta} 
-\beta K(\mid{\bf r}^\alpha_i -{\bf r}^\beta_j\mid)$\cite{gimcdg}.

Eq.~(\ref{eq39}) resembles the partition function of a system of $n$
``species'' of particles, each labeled by an appropriate replica
index and interacting via a two-body interaction which
depends both on particle coordinates $({\bf r}_i,{\bf r}_j)$ and
replica indices $(\alpha,\beta)$.  This system of $n$ species of
particles can be treated in liquid state theory by considering it to be
a $n$-component mixture~\cite{hanmac} and taking the $n \rightarrow 0$
limit in the Ornstein-Zernike equations governing the properties of the
mixture. These equations involve the pair correlation functions
$h^{\alpha\beta}$ and the direct correlation functions
$C^{\alpha\beta}$ of the replicated system.  Assuming replica symmetry
$C^{\alpha\beta} = C^{(1)}\delta_{\alpha\beta} + C^{(2)}(1 -
\delta_{\alpha\beta})$ and $h^{\alpha\beta} =
h^{(1)}\delta_{\alpha\beta} + h^{(2)}(1 - \delta_{\alpha\beta})$.

Ref.~\cite{gimcdg}, obtained $h^{(1)}(\rho,nd)$ and
$h^{(2)}(\rho,nd)$ for the layered vortex system in BSCCO using the HNC
closure approximation.  To estimate the inter-replica interaction
$\beta K(\rho,nd)$, the principal source of disorder was assumed to be
atomic scale pinning centers such as oxygen defects~\cite{chudnovsky}.
(Similar point defects are believed
to act as the principal sources of disorder in the systems
discussed in this paper.)
A model calculation then yields
\begin{equation}
\beta V^{(2)}(\rho,nd) = 
-\beta K(\rho,nd) \simeq -\Gamma^\prime\exp(-\rho^2/\xi^2)\delta_{n,0};
\end{equation}
where $\beta V^{(2)}(\rho,nd) = \beta V^{\alpha\beta}(\rho,nd)$ with
$\alpha \neq \beta$.
$\xi \simeq 15 \AA$ is the coherence length in the $ab$ plane, and
$\Gamma^\prime \approx 10^{-5}\Gamma^2$ for point pinning of strength
$dr_0^2H_c^2/8\pi$, with $d$ the interlayer spacing ($\sim 15\AA$),
$\Gamma = \beta d \Phi^2_0/4 \pi \lambda^2\/$ 
and $\beta=1/k_B T\/$.  Defect densities of the order of $10^{20}$/cm$^3$ 
are assumed in the calculation of the prefactor\cite{gimcdg}.

The calculated correlation functions $C^{(1)}(r)$ and $C^{(2)}(r)$ are
then used as input into an appropriately generalized (replicated)
version\cite{gimcdg} of the density functional 
theory\cite{ry,usreview},
in order to examine the effects of disorder on the
freezing transition.  Applying the replica treatment leads to the
following functional in the $n\rightarrow 0$ limit:
\begin{eqnarray}
\frac{\Delta \Omega}{k_B T}&=&\int d{\bf r}
\left[\rho({\bf r})\ln \frac{\rho({\bf
r})}{\rho_\ell} - \delta \rho({\bf
r})\right] \nonumber \\
&&-\frac{1}{2}\int d{\bf r} \int d{\bf r}^\prime [C^{(1)}(\mid
{\bf r}-{\bf r}^\prime \mid) - C^{(2)}(\mid {\bf r}-{\bf r}^\prime \mid)]
[\rho({\bf r})-\rho_\ell][\rho({\bf
r^\prime})-\rho_\ell] + \ldots.
\label{eq312}
\end{eqnarray}
It is assumed that the density field is the same in all the
replicas ($\rho^\alpha({\bf r}) = \rho({\bf r})$ for all $\alpha$).
The uniform liquid density is $\rho_\ell$. 
This density functional resembles the density functional
of a {\em pure} system
with an {\em effective} direct correlation function given
by
\begin{equation}
C^{eff}(\mid {\bf r} - {\bf r}^\prime \mid) = 
C^{(1)}(\mid {\bf r}-{\bf r}^\prime \mid) - C^{(2)}
(\mid {\bf r}-{\bf r}^\prime \mid),
\end{equation}

In mean-field theory, the freezing transition of the pure system occurs
when the density functional supports periodic solutions
\begin{equation}
\rho({\bf r}) = \sum_{\bf G}\rho_{\bf G}\exp(i{\bf G\cdot r}),
\end{equation}
with a free energy lower than that of the
uniform fluid\cite{ry,usreview}. Here ${\bf G}$ indexes the reciprocal
lattice vectors of the crystal and $\rho_G$ represent the order
parameters of the crystalline state. The properties of the freezing
transition are controlled by $C^{eff}(q)$, the Fourier transform of
$C^{eff}(r)$.  In a one-order parameter approximation, only
$C^{eff}(q)$ evaluated at the wave-vector $q$ corresponding to the
nearest neighbour spacing $a$ {\it i.e.} at $q = q_m = 2\pi/a$ is
required as input. The one-parameter density functional calculation of
Ref.\cite{gimcdg} indicated that melting occured when the $\rho_\ell
C^{eff}(q=q_m)$ attained a value of about 0.8\cite{usreview}. This result
agrees with the phenomenological Hansen-Verlet criterion for the
freezing of a simple two-dimensional liquid\cite{hanmac}:
Freezing occurs when the structure factor 
\begin{equation} 
S(q) = 1/(1- \rho C^{eff}(q)),
\end{equation} evaluated at $q_m$ {\it i.e.} $S(q_m)$
attains a value of about 5.

The following feature of the replicated density functional provides
useful physical insight:
Since C$^{(2)}(q_m) \geq 0$, and C$^{(1)}(q_m)$ is always reduced
(although weakly) in the
presence of disorder, the equilibrium melting line is always
{\em suppressed} by quenched disorder. The analysis of Ref.\cite{gimcdg}
showed that this suppression was very weak at low $H$ but systematically
increased at large $H$, indicating the increased importance of
disorder at high fields.

The semi-analytic approach to the suppression of the MG-DL phase
boundary outlined here uses the following ideas:
\begin{enumerate}
\item The {\em diagonal}
direct correlation function $C^{(1)}(q)$ is very weakly affected by
disorder and can thus be approximated by its value in the
absence of disorder. 
\item The off-diagonal
direct correlation function $C^{(2)}(q)$ is a strong function of disorder
and of $H$, the magnetic field
\item  $C^{(2)}(q)$ is well approximated at
$q=q_m = 2\pi/a$ by its value at $q=0$.  
\end{enumerate}

Since the Hansen-Verlet condition is satisfied along
the melting line, the following is true at melting:
\begin{equation}
\rho_\ell C^{eff} = \rho_\ell(C^{(1)}(q_m) +  C^{(2)}(q_m))\simeq 0.8.
\end{equation}

Now $C^{(2)}$ is a sharply decaying function in real space; its support
is $\xi^2$, which for typical $H \ll H_{c2}$, is far less than $a^2$.
In Fourier space, therefore, its
value at $q = q_{m}$ is close to its value at $q=0$.
We can approximate\cite{hanmac,us} 
\begin{equation}
C^{(2)}(r) \simeq -\beta V^{(2)}(r).
\end{equation}
This scales with temperature as $\Gamma^2$, implying that 
\begin{equation}
\rho_\ell C^{(2)}(q_m) \sim \frac{B}{T^2}.
\end{equation}
Note that $C^{(2)}(q_m)$ increases as $B$ is increased or as $T$ is 
decreased, as is intuitively reasonable.

We turn now to $C^{(1)}(q_m)$. $C^{(1)}(q_m)$ increases with a decrease
in $T$; reducing $T$ increases correlations. The
variation in $C^{(1)}$ is expected to be smooth. We can therefore
write 
\begin{equation}
C^{(1)}(q_m;T - \Delta T,B) =  C^{(1)}(q_m;T,B) - q(B,T)\Delta T,
\end{equation}
where $q(B,T)$~$(q < 0)$ is a smooth function of $B$ and $T$ close to the
melting line.  We are trying to find the effects to first order
of adding $C^{(2)}$, so $(B,T)$ can be replaced by
$(B_m,T_m)$ in Eq. 4.8 above and $C^{(1)}(q_m;B_m,T_m)$ by
its value at freezing for the pure system: $C^{(1)}(q_m;B_m,T_m) 
\simeq 0.8$. The further approximation of neglecting the
$B$ and $T$ dependence of $q$ {\it i.e.} $q(B_m,T_m)\simeq q$,
with $q$ a constant can also be made; at melting this dependence 
should be small provided $a \ll \lambda$.

An approximate expression for the suppression of the melting line
from its value $(B_m,T_m) = (B_m(T),T)$
can now be obtained. Using $q\Delta T_m = pB_m/T_m^2$, 
\begin{equation}
\Delta T_m = \frac{p B_m}{q T_m^2} = m\frac{B_m}{T_m^2} 
= m\frac{B_m(T)}{T^2},
\label{eqshift}
\end{equation}
with $m=p/q$ approximately constant. Here $\Delta T_m$ is
the shift in the melting temperature induced by the
disorder.
This relation predicts a larger suppression of the melting
line at higher fields and lower temperatures, precisely as in the work
of Ref.\cite{gimcdg} and in the experimental data.  Given a
parametrization of the pure melting line, Eq.(\ref{eqshift}) can be used
to estimate the effects of weak disorder on this line.

This result can be combined with results from a calculation of the 
melting line in the pure system to obtain a simple analytic formula 
for the MG-DL phase boundary. At low fields, a Lindemann parameter-based
calculation of this phase boundary yields 
\begin{equation}
B_m(T) = C(T-T_c)^2,
\end{equation}
where T$_c$ is the critical temperature and $C$ is a constant\cite{review1}.
Coupled with Eq.(\ref{eqshift}) above, this yields, in the presence of disorder
\begin{equation}
B^{dis}_m(T) = C(T + \frac{Cm(T-T_c)^2}{T^2} - T_c)^2.
\label{eqvgdl}
\end{equation}

This relation is plotted in Fig. 5 for different values of $m$
together with the melting line in the absence of disorder {\it i.e.}
with $m = 0$. The data shown in Fig. 5 use a T$_c$ value of 7K, as
appropriate for 2H-NbSe$_2$, a prefactor $C$ of 1.15 kG (chosen purely
for illustrative purposes) and values of $Cm= 0,0.2$ and $0.5$.  Note
that the suppression of the pure melting line by disorder is very weak
if $m$ is small. This suppression becomes progressively large as $m$ is
increased, or alternatively, at a lower temperature (larger $H$) for
given $m$. These results are consistent, both qualitatively and
quantitatively with the numerical results of Ref.\cite{gimcdg}. They
are also consistent with the experimental results of Ref. \cite{jsps}
which find that fits of the T$_{p}$ line in disordered samples of
2H-NbSe$_2$ to a Lindemann-type expression are accurate at low field
values but become increasingly inaccurate at larger fields. At large
$H$ the T$_{p}$ line is systematically suppressed to lower $T$ {\it vis
a vis.} the fit, this suppression becoming apparent at lower fields for
more disordered samples.

These results were motivated using a parametrization of vortex lines in
terms of interacting pancake vortices, a description valid for highly
anisotropic layered superconductors in which the c-axis coherence
length is far smaller than the interlayer spacing. How do they
generalize to interacting {\em lines}?  The calculation itself uses two
quantities -- the direct correlation function in the fluid phase for
the pure system, evaluated at a wave-vector equal in magnitude to the
first reciprocal lattice vector of the crystal $G_1$ {\it i.e.} $C(\mid
k_\perp\mid = \mid G_1 \mid,k_z=0)$. This quantity accounts for the
tendency towards ordering at the length-scale of the inter-vortex
spacing.  The calculation which lead to Eq.(\ref{eqshift}) above used
only two properties of this correlation function: (i) the pure system
freezes when the correlation function evaluated at this wave-vector
achieves a particular (universal) value and (ii) the variation of this
quantity upon a reduction of temperature can be parametrized simply,
essentially linearly. Both these features should apply to the case of
interacting lines. Numerical simulations of interacting lines provide
some evidence for similiar quasi-universal attributes of the melting
transition as seen in two and three dimensions, both in simulations and
in experiments\cite{usreview}.

The other quantity required in the calculation, $C^{(2)}$, derives from
properties of the correlation function of the disorder potential
$K(x-x^\prime)$. This is independent of the (line or point) nature of
the model for the flux-line system used.  Thus, it is apparent that the
result obtained in Eqn.(\ref{eqvgdl}), should be a fairly robust one
which does not depend in detail on the fact that it was derived using
results obtained from a model of interacting pancake vortices with only
electromagnetic interactions.

\section{Comparison to Experimental Results}

Section III obtained expressions for the $B-T$ phase boundary
separating BrG and MG phases given H$_{c2}(T)$ and $\lambda$(T). In
Section IV, qualitative and quantitative arguments were provided for
the effects of weak disorder on the MG-DL transition line.
This section compares the predictions of these 
sections with some of the available data on the phase boundaries in the
experimental systems studied in Ref.\cite{mypaper} as well as data on
the high-T$_c$ material Nd$_{2-x}$Ce$_x$CuO$_y$ (NCCO).

We do not attempt fits either to the YBCO or BSCCO data (although
experiments relating to these are discussed in some detail in the
following section), primarily because the results derived above ignore
the effects of thermal renormalization of disorder, which are not
negligible in these materials. (Effectively, these should lead to a
strong $T$ dependence of $\Sigma$, which cannot be treated as a
constant.) We have also not attempted to provide highly
accurate fits to the experimental data in any of the cases below, since
we merely wish to demonstrate that reasonable agreement with data can
be obtained within the theoretical framework described in this paper.

\begin{itemize}

\item {\bf 2H-NbSe$_2$:} 
Fig. 6 plots the experimental data for ($H_{pl},t_{pl}$) in single
crystals of 2H-NbSe$_2$, taken from Fig. 2(a) of Ref.\cite{jsps}
together with a best fit line based on Eqn.(\ref{eqbgvg1}).  As argued
here (also see Ref.\cite{jsps}), this line represents the BrG-MG phase
boundary; t$_{pl}$ is T$_{pl}/T_c$.)  An upper critical field of
H$_{c2}(0) = 46 kG$ is assumed in the normalization of $b$. Assuming
$\kappa \sim 16$, $p \sim 1$ and $c_L \sim 0.22$, yields $\Sigma \sim
60.0$.  The data are represented by an almost straight line over this
field range, with deviations to the tune of about 1kG appearing in the
lower temperature range spanned by the data. The linear behaviour of
B$_{c2}(T)$ assumed in the use of Eq.(\ref{eqbgvg1}), is a feature
of the experimental data in this field and temperature range.

\item {\bf CeRu$_2$:} Data points representing the BrG-MG phase
boundary, extracted from plots of ($H_{pl},T_{pl}$), in Fig. 3(b) of
Ref.\cite{satya1} are shown in Fig. 7, together with a best fit based
on Eqn.(\ref{eqbgvg2}).  An upper critical field of 68kG and a T$_c$ of
6.2 K is assumed in the normalization to aid comparison to experimental
data. Values of $c_L \sim 0.18$, $\kappa \sim 1$ and $p \sim 1$ are
assumed, yielding $\Sigma \sim 0.01$.

\item {\bf Ca$_3$Rh$_4$Sn$_{13}$:} Points extracted from the plots of
($H_{pl},t_{pl}$) in Fig. 3(c) of Ref. \cite{jsps} are shown in Fig.
8, together with best fits based on Eqn.(\ref{eqbgvg2}.  An upper
critical field H$_{c2}$ of 45 kG is assumed in the normalization of the
$y-$axis in the comparison to experimental data.
Values of $c_L \sim 0.18$, $\kappa \sim 1$ and $p \sim 1$ are
assumed, yielding $\Sigma \sim 0.01$.

\item {\bf NCCO:} 
Fig. 1 of Ref.\cite{giller1} illustrates a very successful
phenomenological fit to the experimental data on the BrG-MG phase
boundary in NCCO. This phase boundary was assigned to the locus of
onset points of the fishtail anomaly.  We plot the line obtained using
Eq.\ref{eqbgvg2} with $\Sigma = 0.001$, multiplied by an appropriate
prefactor chosen to allow both curves to overlap as far as possible
together with this fitting form $B_m = B_0(1-(T/T_c)^4)^{3/2}$ in Fig.
9. (The somewhat smaller value of $\Sigma$ here should reflect the
increased role of disorder in these systems, as reflected in the
parameter $p$.) Eq.\ref{eqbgvg2} predicts a marginally smoother and slower
variation of the transition curve than the fitted form, but the
qualitative trends appear to be the same.  As a cautionary note,
the identification of the BrG-MG phase
boundary with the onset curves of the fishtail effect is not
universally accepted.  

\end{itemize}

\section{Discussion of Experiments}

To what extent are either of the two scenarios
outlined in the Introduction supported by the
experiments and the simulations?  To support our
proposal of the second scenario as a {\em generic} one
for type-II superconductors with point pinning
disorder, we take the following approach:  First, we
will show that the situation in which a sliver of
intermediate phase intervenes between the BrG and the
DL phases is actually far more common than earlier
realized. We do this by showing that more recent
experimental data on the high-T$_c$ materials which
initially showed a single melting transition strongly
favours a two-step transition.  Second, we will show
that the first of the two transitions out of the BrG
phase illustrated in Fig. 2, the ``fracturing
transition'' into the MG phase, may not show up at all
in many types of experiments commonly used to probe
phase behaviour, such as measurements of the dc
magnetization.

This second point is illustrated through the following
estimate: If a finite density of unbound dislocations $\rho_d$
enters the sample at the BrG-MG phase boundary, 
the magnetization discontinuity across this boundary should
scale, for small $\rho_d \sim 1/R_d^2$, as
\begin{equation}
\Delta M \sim  \Delta M_0 (a/R_a)^2, 
\end{equation}
where $\Delta M_0$ is the magnetization jump in the pure system at the
melting transition. Even if R$_a/a \sim 30$, the corresponding
induction jump $\Delta B \sim \Delta M$ is of order $10^{-3}M_0$ or
within noise levels in a typical experiment\cite{zeldov,soibel}.

Some of the discussion will focus on the 
peak effect in critical currents seen close to 
the H$_{c2}$ phase
boundary in weakly disordered systems.  
Within a simple Bean model\cite{bean}
for the critical state, the real part of the complex susceptibility,
$\chi '$, obeys
\begin{equation}
\chi '\sim -\beta\frac{j_c }{h_{ac}},
\end{equation}
once the ac field has penetrated fully within the sample. Here $\beta$
is a geometrical  quantity related to the shape of the sample and
h$_{ac}$ is the amplitude of the ac field. Thus, increases in the
critical current density j$_c$ translate to reductions in $\chi '$.
Transport measurements access j$_c$ more directly, although the
presence of Joule heating in the transport measurements is often a
significant factor, particularly in the peak regime where thermal
instabilities are strong.

The observation of an anomaly in magnetic
hysteresis in the high-T$_c$ superconducting oxides has attracted much
attention in the past decade. This anomaly, an {\em increase} in the
width of the magnetization hysteresis curve with field and a concomitant
increase in $j_c$ (following an initial increase and subsequent
decrease due to complete field penetration, the ``first peak'') is
the ``second peak'' or ``fishtail'' anomaly
\cite{kopylov,daeumling,chikumoto}. 
The relation between j$_c$ and the
width of the hysteresis loop follows from:
\begin{equation}
j_c \propto \Delta M = M(H\uparrow) - M(H\downarrow),
\end{equation}
the difference in values of the magnetization on the increasing 
and decreasing branches of the hysteresis loop.
While several explanations have been proposed for this phenomenon, this
behaviour is now believed to be correlated, at least approximately,
with the field-driven transition between two vortex solid phases,
identified by a large body of work as BrG and MG phases\cite{fishtail}.

As pointed out in Ref.\cite{mypaper}, the very structure of the phase
diagram of Fig. 2 mandates the following:  The peak effect in $T$ scans
at intermediate values of $H$ should evolve smoothly and continuously
into behaviour characteristic of the BrG-MG field driven transition,
both at low and high $H$. This is a simple consequence of the absence
of a multicritical point. Insofar as the fishtail effect indicates a
field-driven transition from a BrG phase to the MG phase, signatures of
the fishtail phenomena should connect smoothly to signals of the peak
effect on $T$ scans.  Thus, it is natural to argue for a connection
between fishtail effects (in $H$ scans) and peak effects (in
susceptibility and transport measurements on $T$ scans); both are
phenomena which refer to the {\em same} underlying phase transition out
of the Bragg glass phase.  

What is less clear is the precise connection of the fishtail feature to
the underlying transition. There are at least three distinct
characteristic fields associated with the fishtail feature.  These are
the onset field H$_o$ (also called H$_{min}$), the ``kink'' field H$_k$
(also called H$_3$), the field value at which the magnetization shows a
kink, and the peak field H$_p$\cite{giller2} (also called H$_{sp}$).
The temperature dependences of these features can be quite different.
For example, in YBCO, the onset is most often a monotonically
decreasing function of $T$, while the kink and peak features can show
non-monotonic behaviour. It has been argued that the kink feature most
probably signals the underlying transition\cite{giller2}, although
there does not appear to be universal agreement upon this
point.

\subsection{Low-T$_c$ Systems}

This subsection summarizes features of the data on three low-T$_c$
materials: 2H-NbSe$_2$, CeRu$_2$ and Ca$_3$Rh$_4$Sn$_{13}$. 
Some aspects of this discussion have already appeared
elsewhere\cite{mypaper}, so the discussion here is brief.

\subsubsection{\bf 2H-NbSe$_2$} 
Banerjee {\it et. al.} have reported data on ac susceptibility
measurements on single crystals of 2H-NbSe$_2$ of differing
purity\cite{satya2}. The purest crystal, X, has a T$_c$ of 7.2K,
while a crystal of intermediate purity, Y has a T$_c$ value of
about 7K. A more disordered crystal, Z, has a T$_c$ value of
about 6K. The relative purity of the crystal was inferred from the
values of $j_c/j_{dp}$ for these materials.

In sample X, T$_p$ (where $j_c$ first begins to increase) and T$_{pl}$
(where $j_c$ peaks) are almost coincident.  The data indicate an
extremely sharp peak, with a width comparable to or even smaller than
the width of the superconducting phase transition in zero applied
field\cite{satya2,jsps}. The two-step nature of the transition is
obvious in the sample with intermediate level of disorder, while this
is a prominent feature of the most disordered crystal.  The locus of
T$_p$ and T$_{pl}$ lines, as functions of $H$, almost coincide in X,
are resolvable as separate lines in Y and are clearly visible as
separate features in Z. In sample Z, the T$_p$ and T$_{pl}$ lines
clearly move apart at higher field values, suggesting that their
ultimate fate would be to expand into the broad regime of vortex glass
expected at high field values\cite{mypaper}.

In the most disordered samples of 2H-NbSe$_2$, onset and peak values of
the peak effect are clearly and separately distinguishable down to low
field values, where the T$_p$ line begin to turn around, signalling the
appearance of a reentrant glassy phase\cite{ghosh,jsps,mypaper}.  This
feature suggests that the glass at high fields and the glass at low
fields are {\em smoothly connected}, at least in samples with
intermediate to high levels of disorder\cite{mypaper}. As
Ref.\cite{jsps} indicates, for purer samples, the extreme narrowness of
the peak makes resolution of the features of the intermediate sliver
region very difficult, although the peak itself is obtained to low
field values.

Ravikumar and collaborators see a discontinuity in the dc magnetization
across the peak regime in single crystals of 2H-NbSe$_2$ supporting the
existence of a melting transition.  The experiments could not resolve
whether this feature was connected to T$_p$ or to
T$_{pl}$\cite{ravikumar}.  There is some evidence that the
magnetization anomaly is connected to T$_p$
from the combined magnetization and ac susceptibility
measurements of Saalfrank {\it et. al.}\cite{saalfrank}. These
experiments see a sharp peak in the derivative $dM/dT$, most likely
associated with a step in the magnetization as a function of $T$ at the
location of the T$_p$ line. Saalfrank {\it et. al.} note that the main
peak of $dm/dH$ showed no reentrant behaviour while the peak in the ac
susceptibility shifted to lower temperatures at very low applied field
values, supporting the earlier proposal of a reentrant transition to a
low field glassy phase\cite{ghosh}.   As argued here, magnetization
discontinuities signalling the transition to the liquid should
generically be associated with the T$_{pl}$ line. 

A substantial body of work exists on anomalies associated with the peak
effect regime, particularly in the case of 2H-NbSe$_2$
\cite{satya1,satya2,shobo,satya3,henderson,marley,henderson1}.
Ref.\cite{mypaper} proposed that these anomalies should be primarily
associated with the {\em static} properties of the intervening vortex
glass (equivalently, multi-domain) phase.
This idea is discussed briefly here:  (i) In the
experiments, the transition to the anomalous peak effect regime on
temperature scans from within the BrG phase occurs discontinuously via
an apparent first-order phase transition.  Thus the anomalies seen in
the peak regime {\em cannot} be linked to the dynamics of plastic flow
of a vortex lattice or even quasi-lattice state since the lattice phase
is {\em not} linked smoothly to the intervening sliver phase (ii) the
fact that peak-effect related anomalies in the depinning of a vortex
lattice are {\em only} seen in the peak regime indicates strongly that
a static phase transition may be involved and the solution cannot be
traced to the dynamics alone.  This point is stressed since many $T=0$
simulations of plastic flow (which do not have a phase transition)
appear to be able to reproduce (some but not all) features of the peak
effect phenomenon\cite{jensen}.  Finally, (iii) the
concept of ``softening'' of elastic constants leading to a peak effect
and associated anomalies may well be irrelevant; there appear to be few
strong pre-transitional effects at the fracturing transition and the
peak effect signal in $\chi '$ jumps discontinuously in the
experiments.

\subsubsection{\bf Ca$_3$Rh$_4$Sn$_{13}$}
The Ca$_3$Rh$_4$Sn$_{13}$ system was first studied by Tomy {\it
et.al.}, who drew attention both to the prominent peak effect seen in
this system and the related compound Yb$_3$Rh$_4$Sn$_{13}$ as well as
to the substantial similarities obtained with peak effects in other
superconducting systems, notably UPd$_2$Al$_3$ and
CeRu$_2$\cite{tomy1,tomy2}. Tomy {\it et. al.} obtained H$^\star$, the
field at which the onset of the peak occured both from susceptibility
and dc magnetization experiments.  Plots of H$^\star$ {\it vs.} $T$,
show that the transition curve (which is interpreted here as signalling
the BrG-MG phase transition) appears to connect to the H$_{c2}$ line
only at $T \rightarrow T_c$. The region of fields and temperatures
above H$^\star$ was found to show strongly irreversible behaviour, in
contrast to the behaviour below H$^\star$.

Sarkar {\it et. al.} have studied in some detail the magnetic phase
diagram of this compound through ac susceptibility
measurements\cite{shampa}.  Suceptibility data in this material also
show the characteristic two-step feature exhibited by 2H-NbSe$_2$ at
intermediate fields. At low fields,  the T$_p$ and the T$_{pl}$ lines
come close but do not appear to merge at least until temperatures
fairly close to T$_c$.  The large field data are consistent with the
bending backwards of the T$_{pl}$ line indicating that the intermediate
regime of vortex glass expands out at low temperatures and high fields,
precisely as suggested in Fig. 2.

\subsubsection{\bf CeRu$_2$} 
Early work on single crystals of the cubic Laves phase (C15)
intermetallic superconductor CeRu$_2$ by Huxley {\it et.
al.}\cite{huxley,chaddah} described a sharp reversible to irreversible
transition in the mixed state close to
H$_{c2}$\cite{first}.  The possible relation of these results to a
phase transition into a novel modulated superconducting phase (the FFLO
state\cite{FFLO}) was discussed by these authors and later investigated
extensively by others.  Careful longitudinal and Hall resistivity
measurements by Sato {\it et.  al.} have revealed that the peak effect
survives in this material to far larger fields than the original
measurements indicated\cite{sato}. Fig. 2 of Sato {\it et.  al.}
summarizes the results of their measurements and is to be compared to
Fig. 2 of this paper; the H$_p(T)$ curve in this figure is to be
related to our T$_{pl}(H)$. Note that this line appears to extrapolate
smoothly to T$_c$, although signatures of magnetic hysteresis are not
observed below about 2T.  Sato {\it et. al.} point out that their
observations rule out the FFLO state as a possible explanation of the
PE anomaly. These experiments suggest that pinning is substantially
{\em weakened} at intermediate $H$, although signatures of a peak
effect remain.

Transport measurements on CeRu$_2$ by Dilley {\it et. al.} show a
striking peak effect signal, together with a strong dependence of
measured quantities on the magnitude of the transport
current\cite{dilley}.  Dilley {\it et. al.} note the existence of
strong thermal instabilities in the intermediate peak effect regime,
together with profound hysteresis.  Many of these features are seen in
2H-NbSe$_2$\cite{private}.  Muon-spin rotation studies of
CeRu$_2$\cite{yamashita} have been invoked to suggest that a FFLO
mechanism may be involved in this material. These measurements see a
rapid decrease in the second moment $<\Delta B^2>$ of the field
distribution function across a line in the phase diagram; such a
decrease naively suggests a sharp increase in the penetration depth
$\lambda$.  Yamashita {\it et.  al.} point out that pinning induced
line distortions should generically lead to {\rm broadening} of the
line-width and not to its reduction.  However, a sharp transition to a
multi-domain phase, with {\em short-ranged correlations in the field
direction} would give essentially similar results to those of Yamashita
{\it et. al.}\cite{frozen}.

In a recent illuminating study of the peak effect anomaly in single
crystals of CeRu$_2$, Tenya {\it et. al.} have shown that the peak
effect lies in the {\em irreversible} region of the phase diagram,
counter to some previous work\cite{tenya}.  Their results establish the
following: The onset of the peak, at $T_{pl}$, corresponds to an abrupt
transition from a defective lattice to a regular one.  The state of the
sample in the peak regime is argued to be intermediate between an
ordered structure and a fully amorphous one. These ideas are closely 
related to those presented here.

Work by Banerjee {\it et. al.} shows a remarkable similarity between
the vortex phase diagram in a weakly pinned single crystal of CeRu$_2$
with that of more disordered samples of 2H-NbSe$_2$\cite{satya3,jsps}.
This close similarity  has been used to argue that the origin of peak
effect anomalies in these two materials may well be
common\cite{satya3,jsps}.

\subsection{High-T$_c$ Systems}

This subsection summarizes data relating to peak and fishtail effects
in four high-T$_c$ materials: BSCCO, YBCO, NCCO and (K,Ba)BiO$_3$.  

\subsubsection{\bf YBCO} 
Early torsional oscillator experiments of d'Anna {\it et. al.} measured
complex response in the PE regime of untwinned YBCO\cite{danna1}.  A
prescient comment of d'Anna {\it et.  al.} regarding their data is
particularly noteworthy, for its overlap with the proposals made here.
d'Anna {\it et. al.} commented that {\it
..the melting phenomenon, giving rise to the peak effect, the loss peak
and the kink in resistivity in so-called clean YBCO, is in fact a
complex, two-stage phenomenon..}\cite{danna1}. The first stage was
ascribed to a lattice {\em softening} leading to a change in the
pinning regime, as in conventional approaches to the peak effect, while
the second was associated with a collapse of the shear modulus at
melting and/or a decoupling transition.

In transport measurements in fields upto 10T, d'Anna {\it et. al.}
found two boundaries, between which noise due to vortex motion rose to
a maximum\cite{danna}. The lower boundary was identified with the peak
effect; the noise falls below the limit of resolution at another
boundary, somewhat above the first one. These results are to be
compared to those obtained in similiar transport\cite{shobo} and
susceptibility\cite{satya3} measurements on 2H-NbSe$_2$, which see a
sharp increase in noise amplitudes only in the peak effect regime, with
a discontinuous onset.

Ishida {\it et. al.} have simultaneously measured ac susceptibility and
magnetization  and see a sharp peak effect in the field
range 0.1- 1.5T\cite{ishida}. The peak of the peak effect signal in ac
susceptibility was found to correlate exactly to the location of the
magnetization jump which signals melting.  Ishida {\it et. al.} draw
attention to a small dip feature in $\chi '$ at a temperature T$_s$,
preceding the peak and the associated magnetization discontinuity. This
subtle feature is attributed to a ``synchronization effect between
lattice and pinning sites'', as a consequence of lattice softening.
(The possibility of such synchronization has often been discussed in
the past as a mechanism for the peak effect\cite{campbell}.) The close
similarity of this phenomenology with that discussed for low-T$_c$
systems in Ref.\cite{mypaper} is emphasized here. Ishida {\it et.  al.}
also comment on the enlarged regime in which $\chi '$ shows
non-trivial signatures of the transition in comparison to the
magnetization jump which occurs at a sharply defined temperature.

These results are very simply understood using Fig. 2. We would argue
here for the natural identification T$_s$ = T$_{pl}$.  The width of the
transition region is related to the width of the sliver phase, argued
here to be always finite in scenario (2), while the most prominent
signatures of melting should generically be obtained across a {\em
line}, the MG-DL transition line, which we suggest is the true
remnant of the vortex lattice melting line in the pure system.

Shi {\it et. al.}  have observed a ``giant'' peak effect in ultrapure
crystals of YBCO\cite{shi} at fields ranging from 0-4T\cite{shi}.  This
feature is most prominent at intermediate $H$, reproducing the
phenomenology of the peak effect in 2H-NbSe$_2$. Shi {\it et. al.}
demonstrate that the jump in the magnetization coincides, to within
experimental accuracy, with T$_p$. Shi {\it et. al.} also comment that
peak effect signals can be seen in fields up to 6T, although only by
going to far higher ac drive frequencies (1 MHz). This is consistent
both with the weakening of the effective disorder in the intermediate
field regime and the existence of a sliver phase with finite width.

Rydh, Andersson and Rapp\cite{rydh}, in transport studies of untwinned
YBCO, obtain three regimes of flow:  (1) a low temperature creep regime
(2) an intermediate flow/ creep regime (3) a small regime in which the
resistivity drops as a consequence of the peak effect and (4) a
high-temperature fluid regime. Rydh {\it et. al.} find that the dip in
resistivity is correlated to the melting transition. It is suggested
that the width of region III reflects a distribution of melting
temperatures in the solid and argue that the onset of melting occurs at
the minimum of the resistive dip. In our interpretation, regime III
would be assigned to the multi-domain regime.

Suggestive work by Rassau {\it et. al.} \cite{rassau} studies
magnetization relaxation in the vortex solid phase of YBCO.  Rassau
{\it et. al.} argue that a well-defined region exists below T$_m$ which
can be quantified as a coexistence of solid and liquid phases and
demonstrate that the vortex solid state can be pinned with different
strengths, depending on prior history. The processes which lead to
these different states arise across a narrow temperature region
immediately below T$_m$. This is precisely the phenomenology indicated
by experiments on 2H-NbSe$_2$ and related low-T$_c$ compounds which
show a peak effect; we draw the readers attention to Ref.\cite{satya6}
in this regard. We would argue that the intermediate region interpreted
by Rassau {\it et. al.} as relating to a regime of two-phase
coexistence is our multidomain phase.

Detailed insights into  the vortex phase diagram of untwinned YBCO
comes from the work of Nishizaki {\it et. al.}, who present magnetic
measurements on the vortex lattice in a clean sample \cite{nishizaki}.
At large $H$ and low $T$, the data show a fishtail feature in the
magnetization which narrows, as $T$ is increased, to a bubble-like
feature.  Nishizaki {\it et. al.} discuss the existence of an {\em
anomalous reentrant} behaviour in j$_c$, inferred from their data via a
Bean model-based calculation, for temperatures in the range $68K < T <
74K$. They find that $j_c$ drops very sharply at low fields to almost
unobservable levels, then picks up to form a peak; while hysteresis
decreases and disappears at $T$ = 70K in the intermediate field region
(3.5 T$\leq$ $\mu_0H$ $\leq$ 6T), it reappears again and is maximum
around 10.5T.

Nishizaki {\it et. al} comment that this reentrant magnetization is
seen only in high quality samples with a lower pinning force and
comment that these results indicate that the pinning force for the
untwinned samples is remarkably reduced in the intermediate-field
regime. This comment accords with our earlier discussion of the strong
effects of thermal renormalization of disorder in the intermediate
field, interaction dominated regime. The {\em shape} of the
irreversibility line in the data of Nishizaki {\it et. al.} is
particularly noteworthy in this respect.

Very accurate ac susceptibility measurements using a local Hall probe
have been performed by Billon {\it et. al.}\cite{billon} in the field
and temperature regimes in which Nishizaki {\it et. al.} find
reversible behaviour and a single melting transition. These experiments
find that the critical current is actually finite below the melting
temperature but is extremely small ($\sim 0.4 A/cm^2$), leading to an
irreversible magnetization unresolvable by standard superconducting
quantum interference device magnetometry. These experiments see a peak
effect {\em below} the apparent melting transition, a feature {\em
inaccessible} in conventional SQUID based measurements.  These results
clearly illustrate that signals of a two-step transition in weakly
disordered single crystals of high-T$_c$ materials may be very hard to
access, particularly if discontinuities in $j_c$ or magnetization
across the first transition are small.

Nishizaki {\it et. al.} argue that in the high temperature region above
T$_{cp}$, the BrG-MG transition line (H$^\star(T)$ in the notation of
Nishizaki {\it et. al.}) {\em turns down} and meets the H$_m$(T)
melting line {\em below} H$_{cp}$ for irradiated YBCO. Thus they argue
that the second peak effect just below T$_m$ may be closely related to
enhanced vortex pinning due to vortex lattice softening {\it i.e.} the
conventional peak effect.  The connection of this observation to the
discussion of the phenomenology of the peak effect presented in
Ref.\cite{mypaper} is stressed here, as is the link outlined earlier
between fishtail features and PE features, given the phase diagram of
Fig. 2.  

Nishizaki and collaborators have also reported results on the phase
diagram of untwinned YBCO crystals on varying the Oxygen
stoichiometry.  For fully oxidized YBCO crystals (YBa$_2$Cu$_3$O$_y$, y
$\simeq$ 7, T$_c$ $\simeq$ 87.5K), Nishizaki {\it et. al.} find that
the first order melting transition can be tracked to high fields (upto
30T). For optimally doped YBCO (YBa$_2$Cu$_3$O$_y$, y $\simeq$ 6.92,
T$_c$ $\simeq$ 93K) and slightly underdoped YBCO, it is suggested that
a vortex slush phase can exist between the second-order and first-order
transition lines. This ``slush'' phase intervenes between the glass and
the liquid phases of the vortex system\cite{ssten}. 

How is the slush phase to be understood using the ideas presented
here?  Fig. 2 presents a particularly simple topology for that part of
the phase diagram which involves the MG-DL transition line, in which
the first-order MG-DL transition becomes continous for sufficiently
large $H$. Such a phase diagram does not contain a slush phase.
However, the situation could be more complex.  For example, the
continuous part of the MG-DL transition line obtained at large $H$
could meet the first-order part at a critical end point (CEP).  The
first-order transition above the CEP would then be interpreted as a
liquid-liquid transition, between phases of different densities, with a
further, {\em symmetry-breaking} transition into a glassy phase
occurring at lower $T$ . The phase with a larger average density would
then represent the ``slush'' phase\cite{noteslush}.

Abulafia {\it et. al.}\cite{abulafia} study magnetization
relaxation in YBCO crystals, finding two distinct regimes of relaxation
below and above the peak in the fishtail magnetization. These regimes
were interpreted by these authors as signalling plastic vortex creep in
the vortex lattice state.  This interpretation has been questioned by
Klein {\it et.  al.}\cite{abulafia1} (see also Ref.\cite{abulafia2})
who argue that the fishtail effect marks a transition from a low-field
ordered phase to a high-field glassy structure, an interpretation in
agreement with ours.  It is argued that {\em topological order (in the
glassy phase) can be quenched at large length scales, with diverging
barriers as $j$ goes to zero}\cite{abulafia1}, an idea in agreement
with our suggestion that the multi-domain glass has {\em fixed}
topological order unlike the liquid, but glassy attributes.  With this
interpretation of the data of Abulafia {\it et.  al.} their phase
diagram (Fig. 4 of Ref.  \cite{abulafia}), is very similar to the one
we have proposed. In particular, the melting line and the glass
transition line (if it is identified with the locus of fishtail peak
positions), converge only at $T \rightarrow T_c$.

Recent experiments by Pissas {\it et. al.}\cite{pissas} on single
crystal YBCO samples probe somewhat more disordered samples than those
studied by Nishizaki {\it et. al.}. An abrupt change in slope of the
increasing part of the virgin magnetization curves is seen. This
feature is sharpest at intermediate $T$; it broadens or vanishes at low
$T$ and is hard to discern for $T > 75K$.  This abrupt change of slope,
associated with a curve H$_3(T)$, merges with the second peak feature
H$_{sp}(T)$ at a temperature T$^\star$.  Both the irreversibility line
and the melting line lie {\em above} the second peak line.  The
structure of this phase diagram is to be compared to the ones shown in
Ref.\cite{mypaper} for low-T$_c$ materials in which the irreversibility
line lies above the BrG-MG transition line. The phase behaviour depicted
in Fig. 4 of Pissas {\it et. al.} agrees well with the ideas presented
here as well as the structure of Fig. 2 with the identification of
H$_3$ as the BrG-MG transition line.

To summarize, experiments on untwinned YBCO support the presence of an
intermediate phase which intervenes in equilibrium between BrG and DL
phases, particularly at low fields ($H < 2-4 T)$.  For larger $H$, a
substantially reversible regime is entered, close to the putative
melting transition. In this regime, precision experiments see weak
residual irreversibility as well as a peak effect just below the
melting transition. (SQUID-based magnetization experiments see only a
single transition here.) It is natural to interpret these features as
signalling the continuation of the sliver between high and low-field
ends as well as the profound weakening of disorder effects in the
intermediate $H$, interaction dominated regime, as a consequence of the
averaging of disorder by thermal fluctuations.  We argue that this
weakening of effective disorder is responsible for the {\em apparent}
single-step transition experiments see at intermediate $H$.  At still
higher fields, the fishtail feature splits off from the melting line,
yielding an expanding regime of vortex glass phase. The magnetization
discontinuities associated with this melting line survive to very high
fields ($\sim 30 T$) in the purest samples.  For more disordered
samples, the separation between the BrG-MG and the MG-DL transition
lines is plainly apparent.

\subsubsection{\bf BSCCO}

The earliest data to provide thermodynamic evidence of a sharp melting
transition in single crystals of a high-T$_c$ superconductor were the
Hall probe measurements of Zeldov and collaborators on
BSCCO\cite{zeldov}. These experiments obtained a
discontinuity in the magnetic induction sensed by Hall probes at the
surface of a BSCCO platelet. Practically no hysteresis was seen at any
one probe, while different probes showed slightly different values for
the melting field, indicating a spread of melting temperatures within
the sample.  The temperature driven melting transition in relatively
pure single crystals of BSCCO occurs in a highly reversible regime. As
reported for YBCO, the irreversibility line actually lies {\em within}
the domain of the solid phase for clean crystals.  These experiments
see a single melting transition for weakly disordered systems.  These
and related experiments supported earlier structural evidence, from
neutron scattering\cite{neutron} and muon-spin
rotation\cite{review_muon} for a sharp transition out of a
quasi-lattice phase on increasing both $B$ and $T$.

Very recent magneto-optic measurements by Soibel {\it et. al.} have
reexamined this issue\cite{soibel}.  These experiments see a
global rounding of the transition as a consequence of quenched
disorder, argued by these authors to be due to a {\em broad
distribution of local melting temperatures at scales down to the
mesoscopic scale}\cite{soibel}. These experiments reveal a remarkable
and complex coexistence of fluid and solid domains in the larger
crystals with the properties that while the local transition is sharp,
the global solid-liquid transition is rounded by quenched disorder.
Soibel {\it et. al.} found that the interfacial tension between solid
and fluid phases was very small, indicating that the vortex melting
process was governed to a large extent by disorder.

Recent muon-spin rotation experiments, on three sets of crystals whose
properties range from  overdoped to underdoped, show very clear
indications of a two step transition, as pointed out by the authors.
The first transition on increasing $T$ was associated with a decoupling
transition between the layers. In contrast, in our picture, this would
be the transition into a multi-domain phase.  This identification is
supported by the observation of $\alpha$ values close to unity in this
phase indicating a substantial degree of local order.

Soibel {\it et. al.} see substantial ``supercooling'' across the
first-order melting transition, with relatively small domains of fluid
coexisting with domains of solid. The putative supercooled state then
transforms abruptly into the crystal. We point out here that precisely
such a scenario would be obtained in the context of the phase diagram
of Fig. 2 with the single proviso that the intermediate ``coexistence''
regime which Soibel {\it et. al.} see would be assigned to our proposed
``multidomain'' state, an {\em equilibrium} phase with glassy
properties. This is entirely consistent with our interpretation of the
muon-spin rotation results of Blasius {\it et. al.}. We point out that
from related studies on 2H-NbSe$_2$, it is known that the intermediate
state on field cooling is relatively highly disordered, with
short-ranged correlations resembling those in the liquid. By analogy,
one would expect similar disordered states to be seen in the BSCCO case
on field cooling, precisely as seen\cite{soibel}.

Ooi {\it et. al.} in a study of vortex avalanches in BSCCO through
local magnetization and local permeability measurements, see a stepwise
expulsion of vortices in a distinct temperature regime {\em below} the
first-order transition on decreasing temperature scans\cite{ooi1}.
This expulsion ceases abruptly below a second, lower temperature called
by these authors as T$_d$. In between T$_M$ and T$_d$, the experiments
see broad-band noise with a power law spectrum. This
phenomenology precisely reproduces related noise data in the peak
effect regime of 2H-NbSe$_2$ and related systems.

Ooi {\it et. al.} also report a possible single step transition for
weakly disordered samples and a {\em temperature-dependent} peak
effect, which they associate with inhomogeneities in the melting field,
for more disordered samples\cite{ooi2}.  These results, presented as
schematics as Fig. 5 of Ooi {\it et. al.} have interesting parallels
with the phase diagrams presented here. We would argue that the regime
that Ooi {\it et. al.} assign to inhomogeneities in the melting field
should be attributed instead to a genuine thermodynamic phase, the
sliver of disordered MG phase, which intervenes between the BrG and DL
phases. 

A simulation by Sugano {\it et. al.} finds clear evidence for a
two-stage melting of the disordered vortex-line lattice in
BSCCO\cite{sugano}. These authors find that the low-field melting
transition out of the Bragg glass occurs always via an intermediate
``soft'' glass phase; their phase diagram (Fig. 3 of
Ref.\cite{sugano})  is to be compared to the one shown in Fig. 2. Their
proposal that the glassy phase is subdivided into strongly pinned and
weakly pinned regimes has interesting overlaps with our suggestion of
the increased importance of non-equilibrium effects at low $T$.

\subsubsection{\bf NCCO}

The onset of the fishtail peak in magnetization measurements on a
Nd$_{1.85}$Ce$_{0.15}$ CuO$_{4-\delta}$ crystal, a layered cuprate
system with substantial anisotropy and a relatively low T$_c$ ($\sim$
23 K) has been probed via local magnetization measurements using an
array of Hall probes\cite{giller1}. These experiments (in particular
the phase diagram of Fig. 1 of Ref.\cite{giller1}), show an onset field
for the peak effect H$_o$ which is always distinct from the
irreversibility line separating an ``entangled solid'' phase from the
vortex liquid phase. This phase diagram supports the ideas presented
here, in particular the contention that a single sharp melting
transition out of the BrG phase is {\em not} generic (as scenario (1)
would suggest) and provides strong support for scenario (2).

\subsubsection{\bf (K,Ba)BiO$_3$}
Small angle neutron scattering studies of the isotropic high-T$_c$
material (K,Ba)BiO$_3$ (T$_c \sim 30K$) suggest a phase diagram which
is very close to the one displayed in Fig. 2 of this paper\cite{klein}.
In this system, as in the NCCO system discussed previously, the
transition line between the quasilattice and the glassy state always
lies {\em below} the transition line separating the glass from the
liquid.  Klein and collaborators comment that the SANS data in this
system support the connection of the fishtail effect with the
transition to a glass out of the quasi-lattice BrG phase\cite{klein}.
In addition, another interesting feature of the data is the relatively
smooth decrease in the intensity of a Bragg peak on field scans, while
the width of the peak remains virtually constant, suggesting a
relatively high degree of local correlations in the putative vortex
glass phase, as we would expect for a multi-domain solid.

\section{Conclusions}

This paper has presented arguments in favour of a
generic phase diagram for type-II superconductors with
quenched point pinning disorder.  This phase diagram,
Fig. 2, differs from others proposed earlier.  We
suggest that the relatively ordered BrG phase {\em
generically} transforms into an intermediate glassy
state with solid-like correlations out to relatively
large length scales rather than directly into a
liquid.  We have pointed out that many experiments
are, in fact, consistent with this proposal and
discussed how signals of the first transition, from
the BrG to the MG phase, may often be hard to detect.
We have also presented a simple analytic
parametrization for the BrG-MG phase boundary, drawing
on earlier work\cite{giam,giam1} as well as provided
an analytical expression for the MG-DL phase boundary.
These are consistent with the experimental data.

Hypothesizing a multi-domain state in the intermediate-$H$ regime is
consistent with the experimental data.  This suggestion rationalizes
the association of thermodynamic melting with the MG-DL transition
line; for a prior suggestion in this regard see Ref.\cite{kiervin}.  In
addition, given the expectation that the effects of disorder increase
as $H$ increases, it indicates a physical mechanism for the occurence
of a critical point on the first-order MG-DL transition line.

Related theoretical and simulation work which bears on ideas proposed
here include work by Feinberg\cite{feinberg}, Ikeda\cite{ikeda} and van
Otterlo {\it et. al.}.  Ikeda has suggested that a vortex glass
instability may accompany a first-order solidification transition in
the presence of weak point pinning disorder\cite{ikeda}.  Feinberg
has argued that melting of the Bragg glass may be to an intermediate
glassy phase, {\it via} a reentrance of single particle
pinning\cite{feinberg}.  van Otterlo {\it et. al.} point out that their
simulation work cannot rule out the possibility of a sliver of glassy
phase always intervening between ordered and fluid phases, as in the
phase diagram of Fig. 2.  Several authors have commented \cite{klein},
on the relation of these simulation results to their experimental
data.  Some of the ideas presented here also have overlaps with recent
work by Kierfeld and Vinokur\cite{kiervin}.  A recent comprehensive
survey of the status of vortex glass phases\cite{natterman1} makes much
the same points as we do regarding the absence of a true phase-coherent
MG phase as envisaged in the original proposal\cite{fifi}.  Several of
the ideas presented here draw from extensive work on peak effect
phenomena in 2H-NbSe$_2$ and related materials, summarized in
Ref.\cite{jsps}.

Many of the proposals presented here are, in fact, experimentally
testable. One is our proposal of a multi-domain phase.  Experiments
which probe local correlations should be able to access such
structure.  We have suggested that in the regime where the sliver of
intervening glassy phase narrows, structural correlations in this
intermediate phase should become large.  This proposal is testable both
in simulations and in experiments.

The second is our identification of the first of the two phase
transitions out of the BrG phase on increasing $T$ as a thermodynamic
phase transition.  Signals of this phase transition in the form of
entropy jumps or singularities in the specific heat will be extremely
small, since they reflect ordering at the scale of units of 10$^4$
vortex lines or more.  However, it remains to be seen whether
high-precision experiments might be able to resolve our suggestion of
{\em two} equilibrium, thermodynamic phase transitions generically
separating the low temperature Bragg glass phase from the disordered
liquid. Such experiments would provide crucial evidence in favour of
the ideas presented here.

Perhaps the single most novel proposal of this paper is the conjecture
that the low-$T$, quasi-lattice Bragg glass phase in {\em all}
superconductors should generically melt into an intermediate glassy
phase before finally transforming into a liquid\cite{menon1}.  This
possibility violates none of what is known about the phenomenology of
the experimental data, nor the simulations nor available theoretical
evidence. Further tests of the ideas presented here as well as
first-principles calculations of the phase diagram of Fig. 2 would be
very welcome.

\section{Acknowledgements} 
I thank my coauthors on Ref.\cite{mypaper} -- S.S. Banerjee, T.V.
Chandrasekhar Rao, A.K. Grover, M.J. Higgins, P.K. Mishra, D. Pal, S.
Ramakrishnan, G. Ravikumar, V.C. Sahni, S. Sarkar and C.V. Tomy.
I thank Deepak Dhar and G.  Ravikumar for useful discussions.  I am
grateful to C. Dasgupta for a critical reading of the manuscript and
for many clarifying discussions. Useful conversations with S. Sondhi
and T. Nattermann are acknowledged and David Nelson is thanked for a
clarifying discussion regarding hexatic vortex glasses. This work evolved
out of many discussions with S.  Bhattacharya -- he exerted a profound
influence in shaping the ideas presented here. The author is partially
supported by a DST (India) Fast Track Fellowship for Young Scientists.

\begin{figure}[htp]
\vskip 10truecm
{\includegraphics{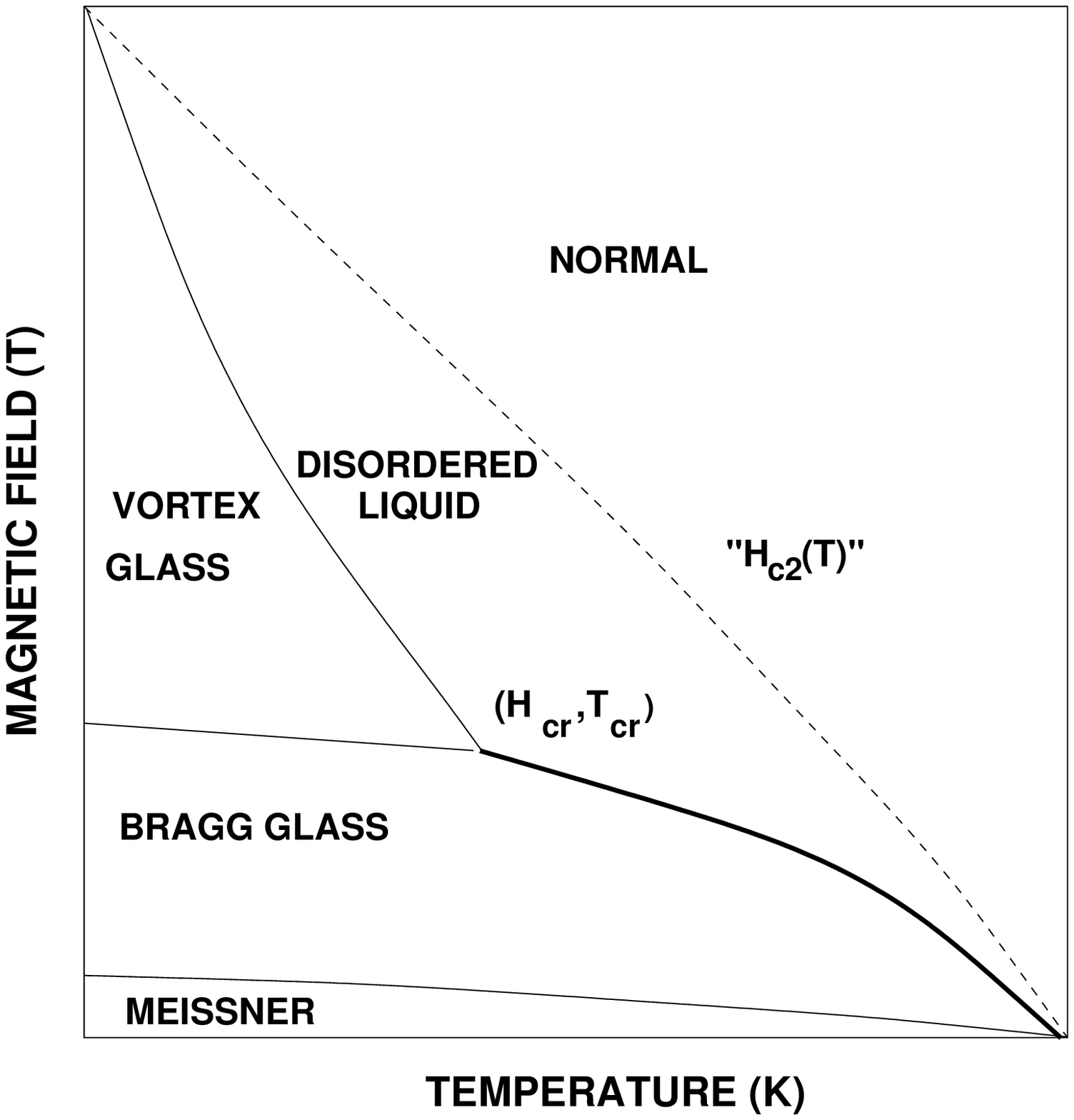}}
\vskip 1truecm
\caption{The current view of the phase diagram of type-II
superconductors\protect\cite{giam1,young,gammel,natterman1,vinokur,kiervin},
incorporating the effects of quenched point pinning disorder and thermal
fluctuations. In addition to Meissner and normal phases, this
phase diagram subdivides the mixed phase into three phases -- the Bragg
glass (BrG), the vortex glass (VG) and the disordered liquid (DL).  The
BrG melts directly into the DL phase through a first-order melting
transition on $T$ scans at intermediate $H$.  On $H$ scans at fixed low
$T$, the BrG phase transforms discontinuously\protect\cite{gaifullin,vanderbeek}
into the disordered VG phase.  The VG phase is understood to transform
via a continous phase transition, with the exponents linked by
relations calculated in Ref.\protect\cite{fifi}, into the DL phase. This line of
continuous phase transitions meets the BrG phase boundary at a
``multicritical'' point, labelled $(H_{cr},T_{cr})$ in the figure. The
first-order direct transition from BrG to DL may persist beyond the
putative (H$_{cr},T_{cr}$)\protect\cite{lopez,nishizaki,kiervin}
terminating in a critical point (not shown in this figure, see
Refs.\protect\cite{kiervin,natterman1}). The line which separates the DL from
the normal phase is a crossover line and not a true phase transition.
This phase diagram does not show the small regime of reentrant glassy
phase expected at low field values; for a phase diagram which includes
this see Ref.\protect\cite{natterman1}.
}
\label{Fig1}
\end{figure}
\newpage
\begin{figure}[htp]
{\includegraphics{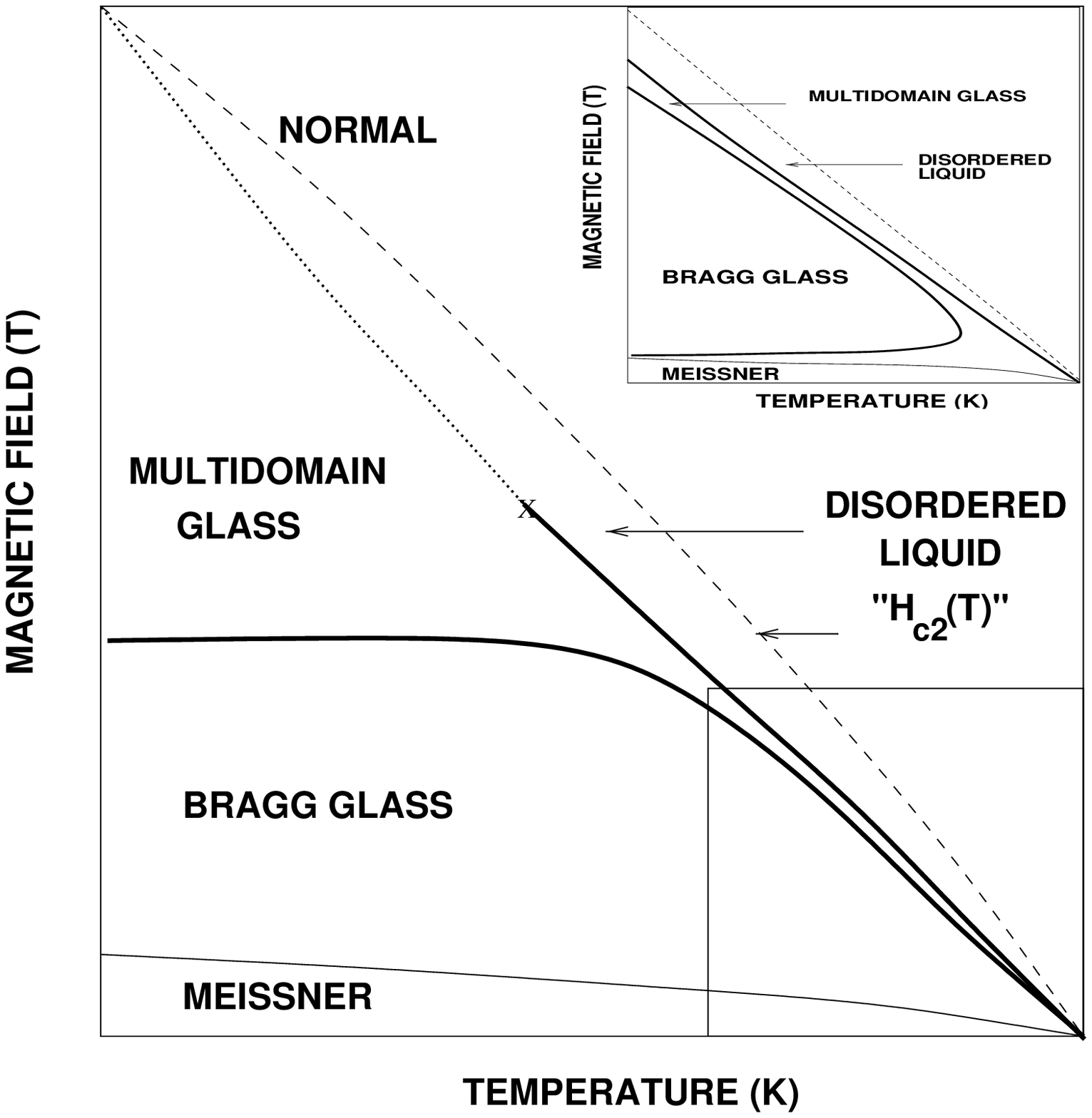}}
\caption
{
Proposed universal phase diagram for type-II superconductors
incorporating the effects of thermal fluctuations and quenched random
point pinning. The term multi-domain glass (MG) is used here in
preference to ``vortex glass''; for  a discussion of this point and of
the properties of this phase, see text.  The other phases are
described in the caption of Fig. 1.  Note that the MG phase intrudes
between BrG and DL phases everywhere in the phase diagram.  At
intermediate $H$ where interactions dominate, the MG phase is confined
to a slim sliver but broadens out again for sufficiently low $H$. The
putative ``multicritical point'' (H$_{cr},T_{cr})$ of Fig. 1 is
identified with the location where the BrG-MG and MG-DL transition line
first approach so closely that they cannot be separately resolved.  The inset
to the figure expands the boxed region shown in the main panel at low
fields and temperature values close to T$_c(0)$ and illustrates the
reentrant nature of the BrG-MG phase boundary at low fields.  The
transition line separating the MG from the DL phase  is first order at
low fields but terminates at a tricritical point, where it meets a 
line of (equilibrium) glass transitions. The transition out
of the BrG phase is generically first-order. 
}
\label{Fig2}
\end{figure}

\newpage

\begin{figure}[htp]
{\includegraphics{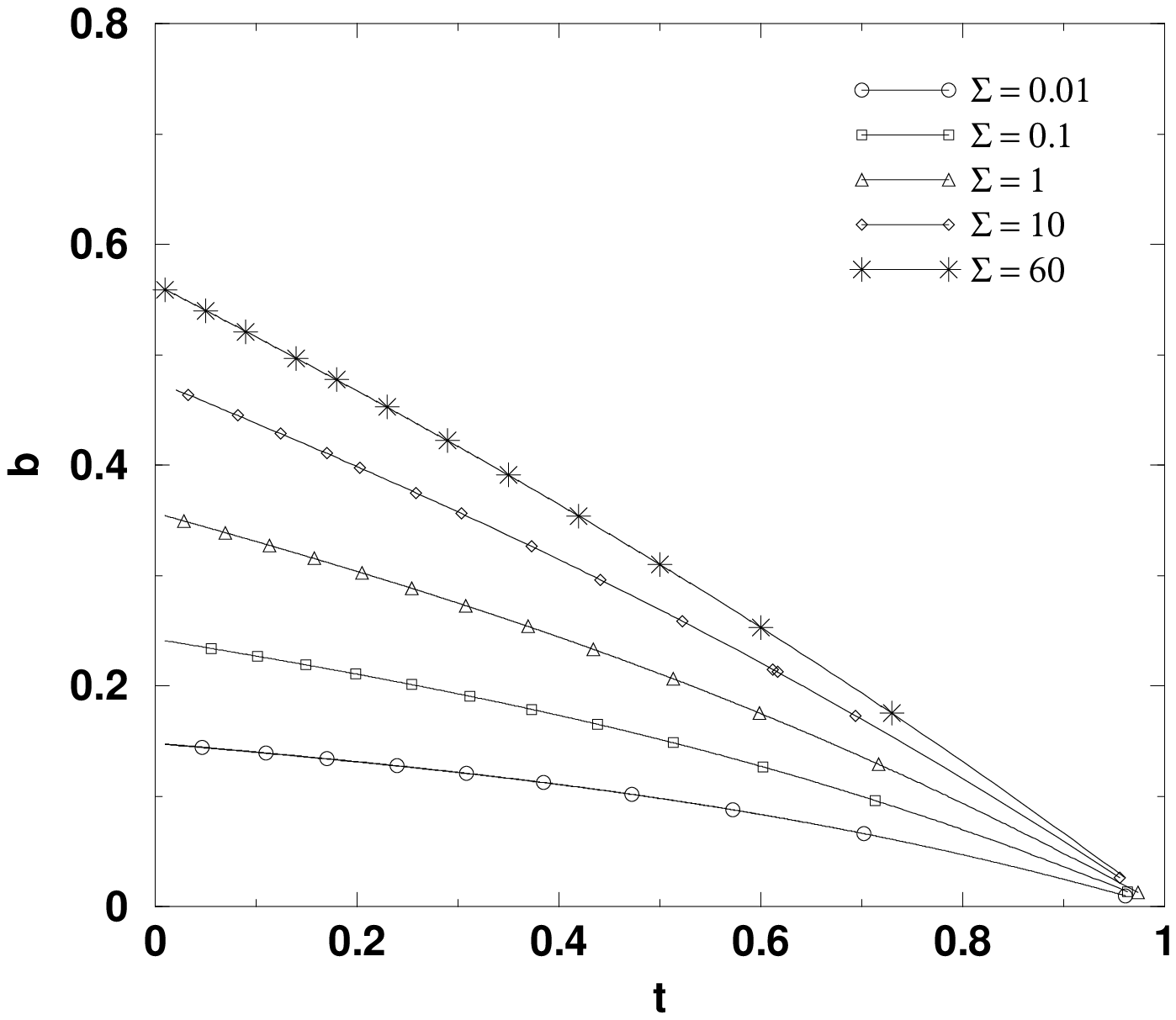}}
\caption
{
Plot of Eq.~\ref{eqbgvg1}, the expression for the BrG-MG phase boundary
in the $b,t$ plane ($b$ and $t$ are the reduced field and temperature
$H/H_{c2}$ and $T/T_c$ respectively), derived in Sec. III, for
different values of $\Sigma$, in the range $0.01 < \Sigma < 60$ (see
text). $\Sigma$ is a phenomenological fitting parameter involving the
Lindemann parameter and fundamental constants.
}
\label{Fig3}
\end{figure}

\begin{figure}[htp]
\vskip 7.5truecm
{\includegraphics{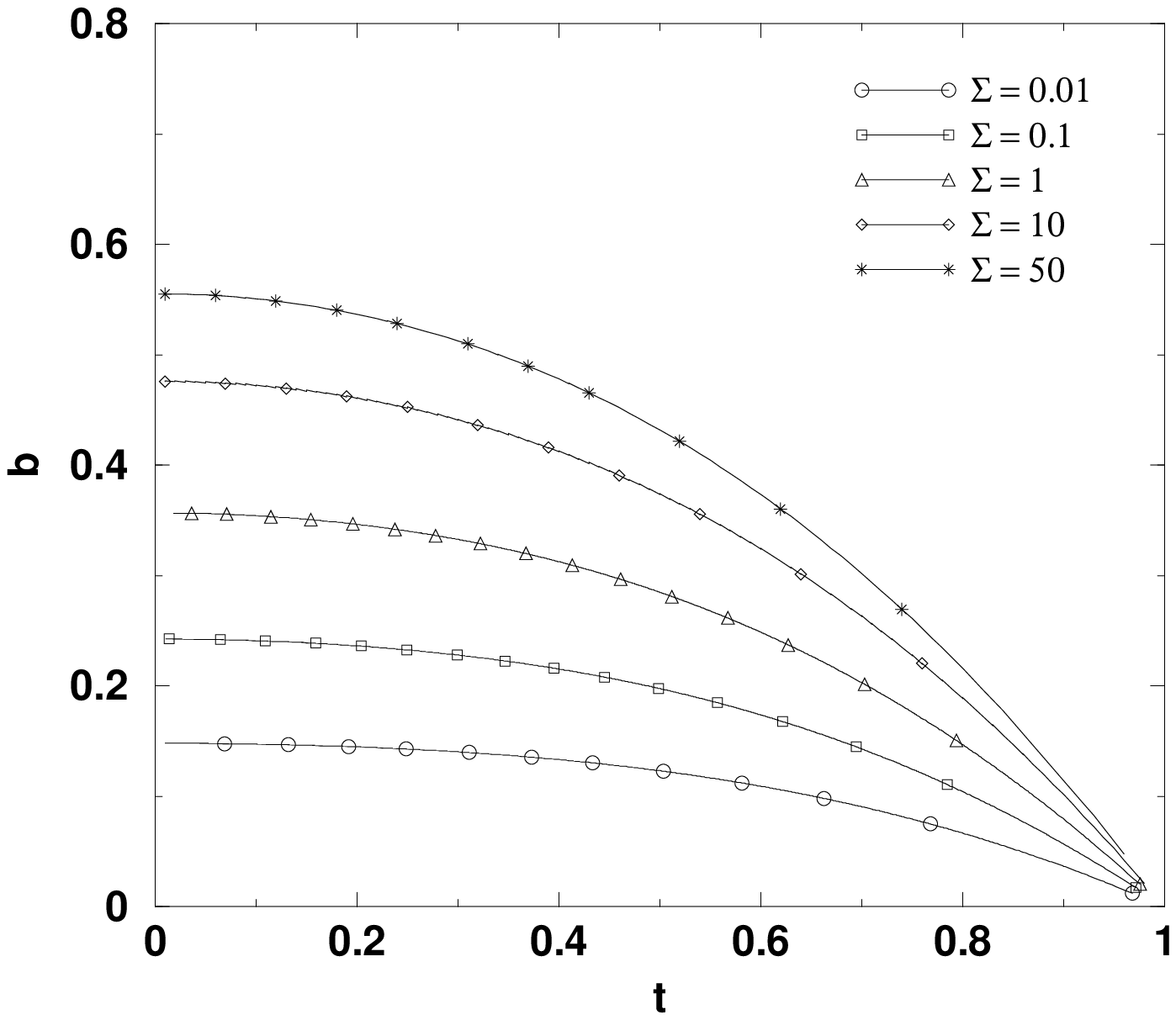}}
\caption
{
Plot of Eq.~\ref{eqbgvg2}, the expression for the BrG-MG phase boundary
in the $b,t$ plane ($b$ and $t$ are the reduced field and temperature
$H/H_{c2}$ and $T/T_c$ respectively), derived in Sec. III, for
different values of $\Sigma$, in the range $0.01 < \Sigma < 60$ 
(see text). $\Sigma$ is a phenomenological fitting parameter involving 
the Lindemann parameter and fundamental constants.
}
\label{Fig4}
\end{figure}

\newpage

\begin{figure}[htp]
{\includegraphics{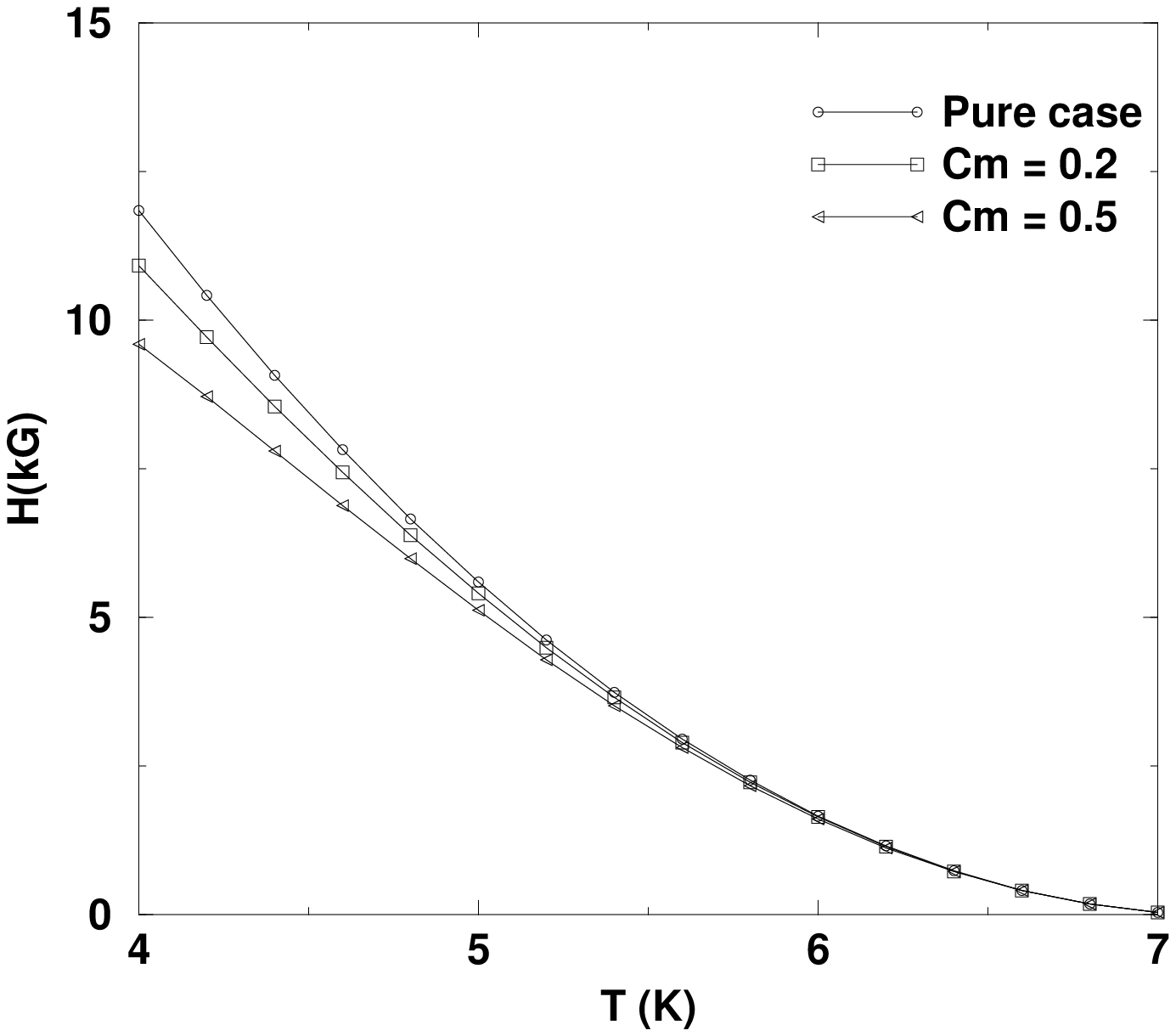}}
\caption
{
Plot of Eq.~\ref{eqvgdl}, the expression for the
MG-DL phase boundary incorporating the effects
of quenched disorder as described in the text, derived in
Sec. IV.  The data use a T$_c$ value of 7K, as
appropriate for 2H-NbSe$_2$, a prefactor $C$ of 1.15 kG (chosen
purely for illustrative purposes) and values of $Cm=
0,0.3$ and $0.5$. Note that the pure melting line is increasing
suppressed  by quenched disorder, this suppression becoming
larger as the applied field $H$ is increased, in agreement with
the predictions of Ref.\protect\cite{gimcdg} and experiments. 
}
\label{Fig5}
\end{figure}

\begin{figure}[htp]
\vskip 7.7truecm
{\includegraphics{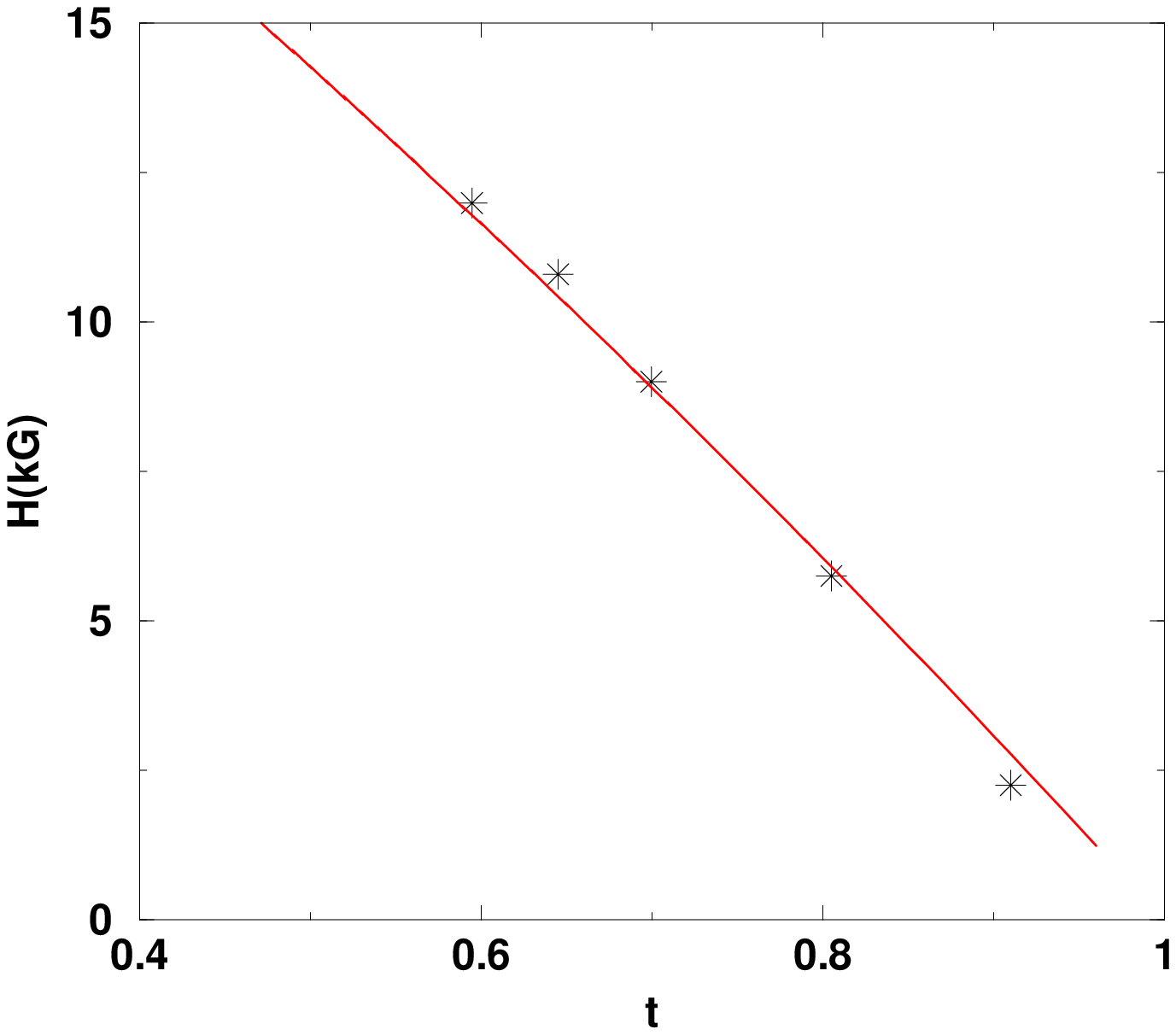}}
\caption
{
Plot of the experimental data for ($H_{pl},t_{pl}$) in single
crystals of 2H-NbSe$_2$ (points taken from Fig. 2(a) of 
Ref.\protect\cite{jsps}) together with a best fit line based on
Eqn.(\ref{eqbgvg1}). Here t$_{pl}$ is T$_{pl}/T_c$.  An upper
critical field of H$_{c2}(0) = 46 kG$ is assumed in the normalization
of $b$ and $\Sigma = 60$ is used; see text.  
}
\label{Fig6}
\end{figure}

\newpage

\begin{figure}[htp]
{\includegraphics{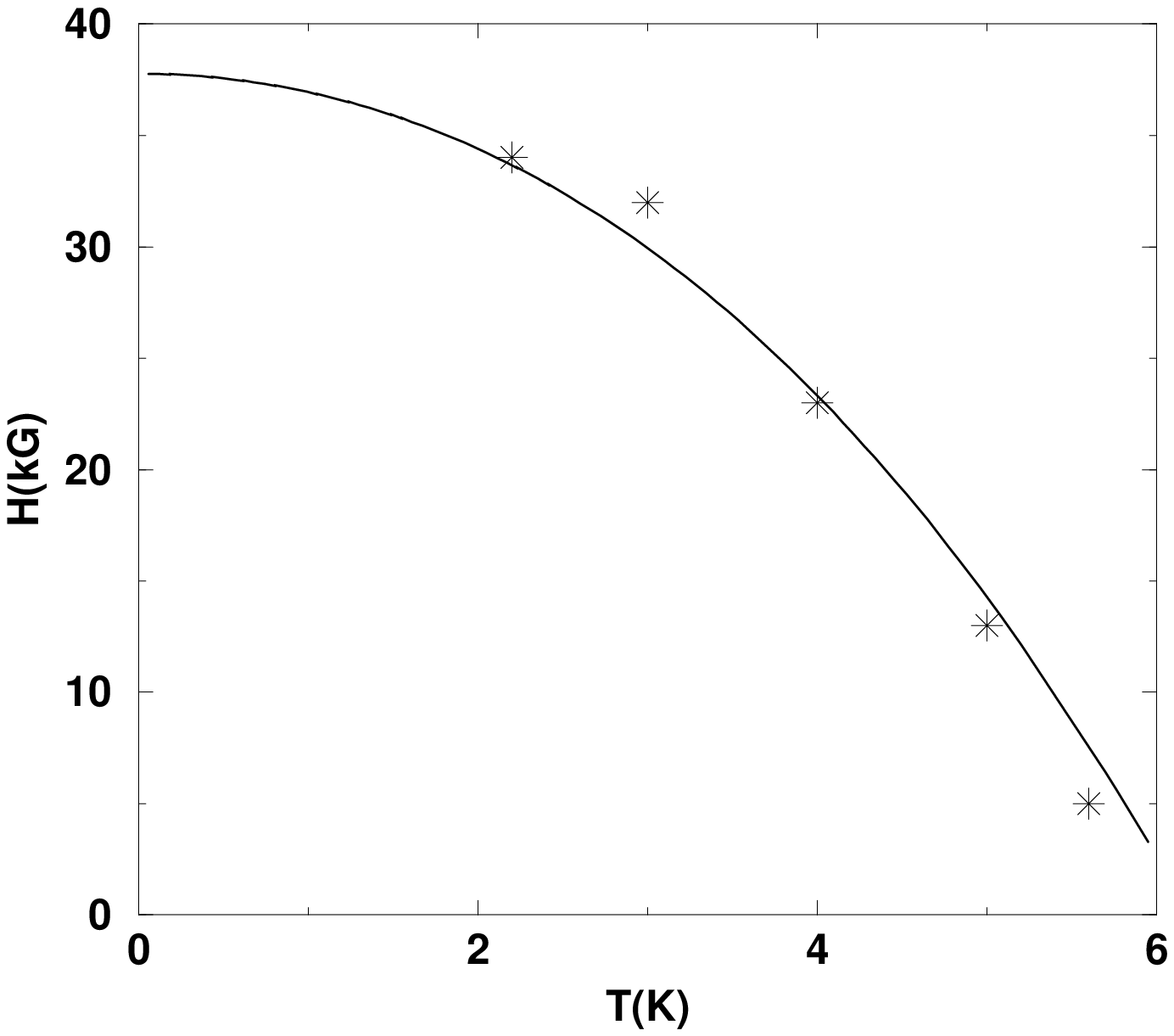}}
\caption
{
Data points representing the BrG-MG phase
boundary in CeRu$_2$ (points taken from plots of ($H_{pl},T_{pl}$)
in Fig. 3(b) of Ref.\protect\cite{satya1}), together with a best fit based
on Eqn.(\ref{eqbgvg2}).  An upper critical field of 68kG and a T$_c$ of
6.2 K is assumed in the normalization of the $y-$axis. A value of
$\Sigma = 0.01$ is used; see text.
}
\label{Fig7}
\end{figure}

\begin{figure}[htp]
\vskip 8truecm
{\includegraphics{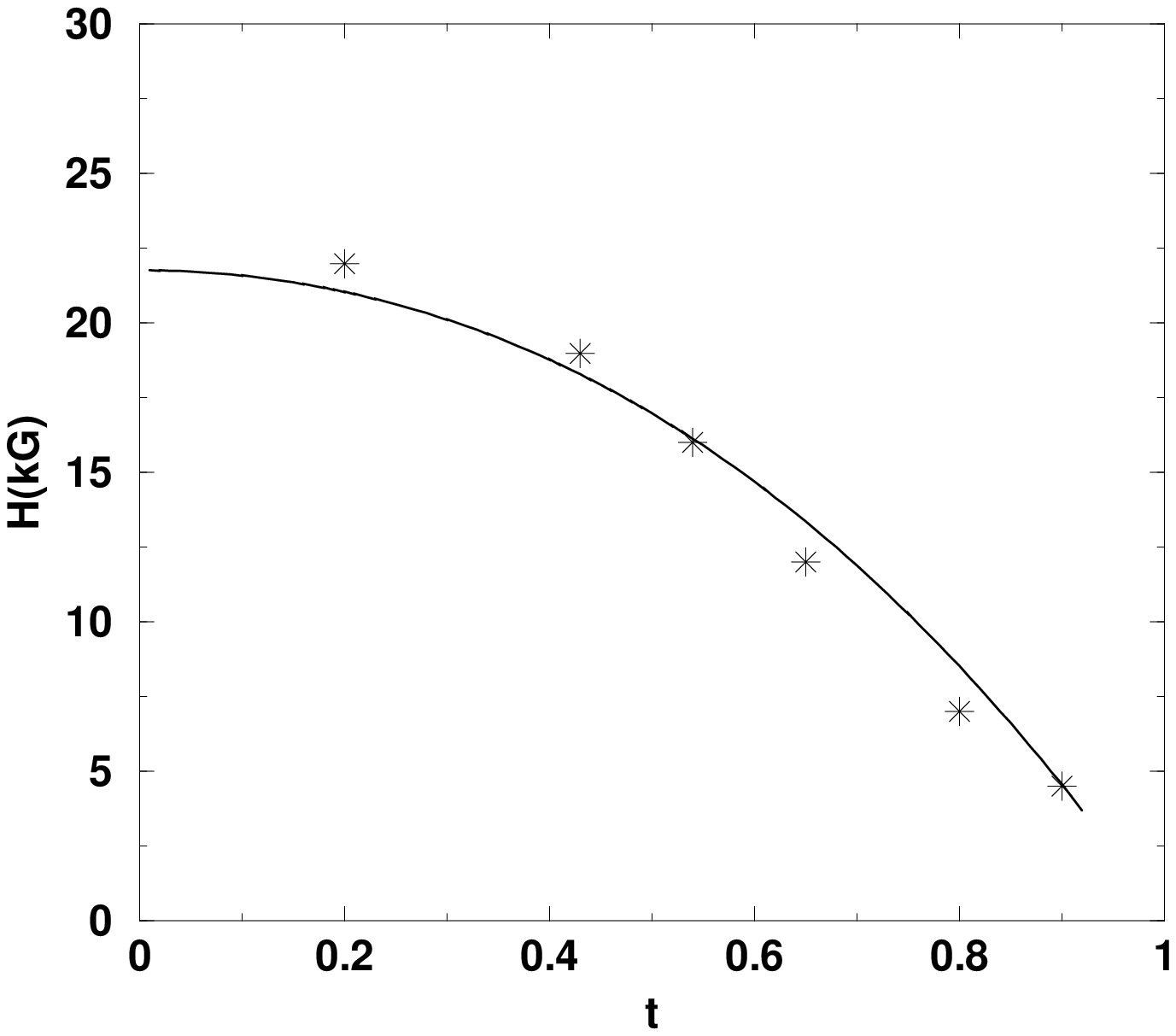}}
\caption
{
Data points obtained from plots of
$H_{pl},t_{pl}$ in Ca$_3$Rh$_4$Sn$_{13}$ (from
Fig. 3(c) of Ref. \protect\cite{jsps}) are shown 
together with best fits based on Eqn.(\ref{eqbgvg1}.  An upper
critical field H$_{c2}$ of 45 kG is assumed in the normalization 
of the $y-$axis in the comparison to experimental data.  A value of
$\Sigma = 0.01$ is used; see text.
}
\label{Fig8}
\end{figure}

\newpage

\begin{figure}[htp]
{\includegraphics{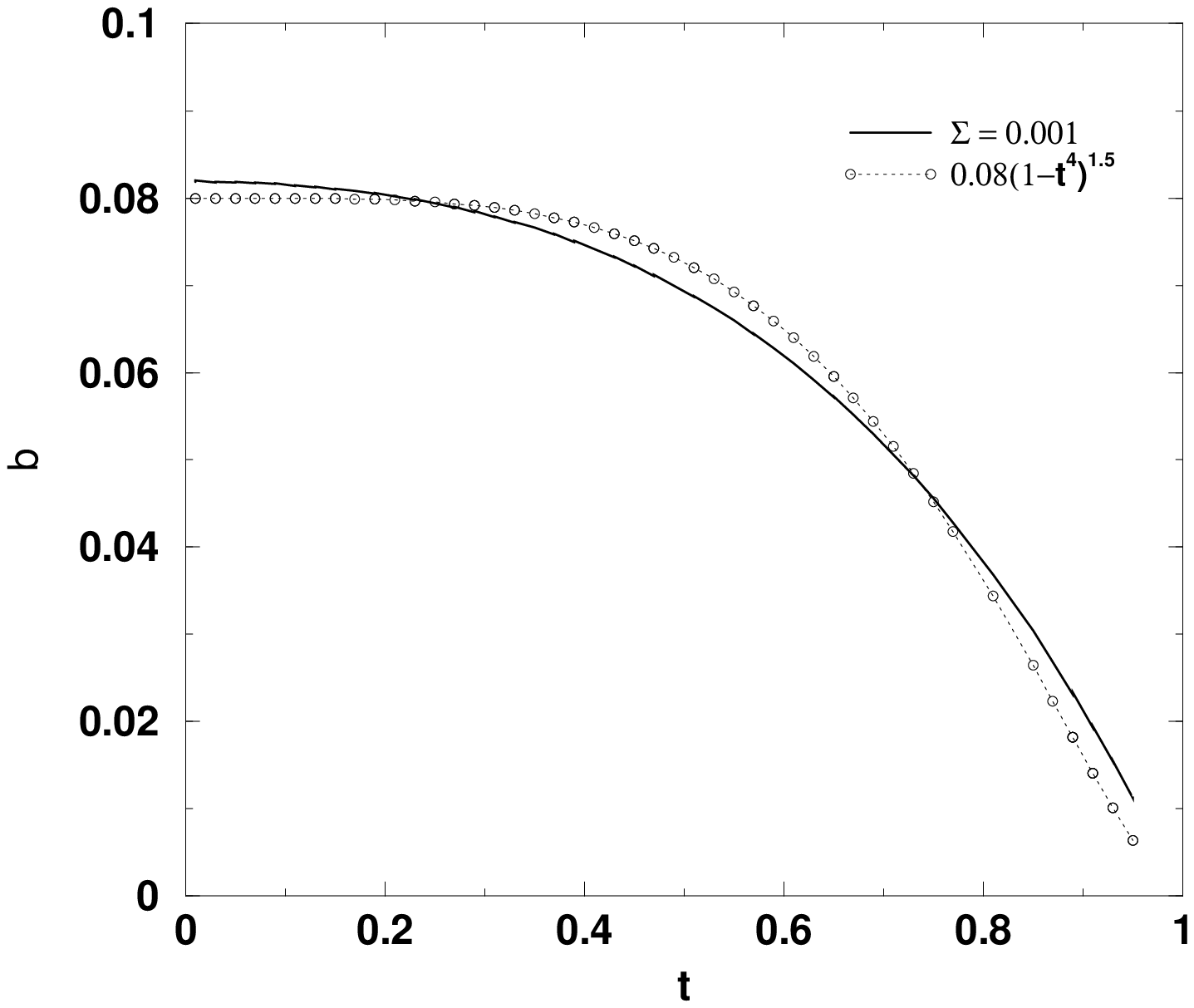}}
\caption
{
The solid line is the solution of Eq.\ref{eqbgvg2} with 
$\Sigma = 0.001$.  The dotted line is the fitting form 
$B_m = B_0(1-(T/T_c)^4)^{3/2}$,
which provides an accurate fit to the onset value of the fishtail
effect in NCCO\protect\cite{giller1}. The prefactors are chosen to
allow the two curves to overlap as extensively as possible. 
}
\label{Fig9}
\end{figure}


\begin{thebibliography}{99}
\bibitem{footnotedis} The dominant source of disorder in these crystals 
is localized pinning sites such as vacancies or interstitials in the 
host crystal.
\bibitem{typedis} We will thus discuss data on systems where
pinning by extended defects such as lattice dislocations, 
grain boundaries, twins or ion tracks is absent.
\bibitem{clarifying} A multi-domain glass in which the
U(1) symmetry of the complex superconducting order parameter $\Psi$
was spontaneously broken would then be equivalent to a vortex glass.
\bibitem{some} Some of the possibilities for the topologies of the
transition lines in such a case
are illustrated in Ref.\cite{gammel}.
\bibitem{fishtail}
In BSCCO, the fishtail effect is a spectacular and sharp feature,
immediately suggesting a link to an underlying thermodynamic phase
transition\cite{khaykovich}. In YBCO and less anisotropic high-T$_c$
materials, the onset of the fishtail feature is more gradual and the
association to an underlying equilibrium transition more indirect.
\bibitem{first} The possibility of a first-order transition in the 
mixed phase was suggested, given the existence of slight hysteretic 
effects, a mild discontinuity in the specific heat and somewhat more 
unequivocal thermo-mechanical signals.  However, no magnetization 
discontinuity was seen.
\end{thebibliography}
\end{document}